\documentclass[reprint,superscriptaddress,prb,aps]{revtex4-2}

\usepackage{mathtools}
\usepackage{amsmath}
\usepackage{amsthm}
\usepackage{amssymb}
\usepackage{graphicx}
\usepackage{bm}
\usepackage{siunitx}
\usepackage[colorlinks,linkcolor=blue, urlcolor=blue,citecolor=blue,bookmarks=false]{hyperref}

\begin{document}

\title{Intra-unit-cell resolved intertwining of multi-\textit{Q} charge and spin textures\\in an itinerant skyrmion magnet}

\author{Christopher J. Butler}
\email{christopher.butler@riken.jp}
\affiliation{RIKEN Center for Emergent Matter Science, Wako, Saitama 351-0198, Japan}

\author{Katsuki Nihongi}
\affiliation{RIKEN Center for Emergent Matter Science, Wako, Saitama 351-0198, Japan}

\author{Haruto Yoshimochi}
\affiliation{RIKEN Center for Emergent Matter Science, Wako, Saitama 351-0198, Japan}

\author{Nguyen Duy Khanh}
\affiliation{Department of Applied Physics and Quantum-Phase Electronics Center (QPEC), University of Tokyo, Tokyo 113-8656, Japan}

\author{\\Rina Takagi}
\affiliation{Institute for Solid State Physics, University of Tokyo, Chiba 277-8581, Japan}

\author{Tetsuo Hanaguri}
\email{hanaguri@riken.jp}
\affiliation{RIKEN Center for Emergent Matter Science, Wako, Saitama 351-0198, Japan}

\author{Shinichiro Seki}
\affiliation{Department of Applied Physics and Quantum-Phase Electronics Center (QPEC), University of Tokyo, Tokyo 113-8656, Japan}
\affiliation{Research Center for Advanced Science and Technology, University of Tokyo, Tokyo 153-8904, Japan}

\begin{abstract}

The mechanisms stabilizing non-collinear magnetism in centrosymmetric crystals remain unclear, but likely involve spin-spin interactions mediated by itinerant electrons, such as the RKKY interaction. Finite-$Q$ magnetic order may then be accompanied by electronic modulations that are observable using a scanning tunneling microscope. In five successive magnetic phases of GdRu$_{2}$Ge$_{2}$, including two nano-scale skyrmion crystal phases, we show that multi-$Q$ magnetism among Gd 4$f$ spins entails a corresponding multi-$Q$ texture among the Ru 4$d$ orbitals that contribute itinerant electron bands. With atomically-resolved images of each electronic texture's motif, and a simple numerical modeling scheme drawing on the underlying spin structures, we infer their key relationship: The alignment between nearest-neighbor Gd spins tightly correlates with the local density-of-states of the Ru 4$d$ orbitals on the two bond-centered sublattices of the Gd square net. These analyses offer a microscopic view of the atomic-scale intertwining of charge and spin degrees-of-freedom in non-collinear itinerant magnets.

\end{abstract}

\maketitle

\section*{Introduction}

Skyrmions are topologically stable winding configurations of a vector field such as the lattice of spins in a magnet. A widely-known form of spin-spin interaction that can stabilize magnetic skyrmions is the Dzyaloshinskii-Moriya interaction, which comes into effect in chiral magnets, and at heterointerfaces and surfaces where inversion symmetry is broken \cite{Dzyaloshinskii1958,Moriya1960,Muhlbauer2009,Yu2010,Heinze2011}. But in recent years, the smallest known skyrmions have instead been found in inversion-symmetric crystals where the Dzyaloshinskii-Moriya interaction is absent \cite{Khanh2015,Kurumaji2019,Hirschberger2019,Ishiwata2020,Gao2020,Takagi2022,Khanh2022}, such as the centrosymmetric skyrmion magnets Gd$_{2}$PdSi$_{3}$~\cite{Kurumaji2019}, GdRu$_{2}$Si$_{2}$~\cite{Khanh2015}, and EuAl$_{4}$~\cite{Takagi2022}. Here it is less clear what the mechanisms stabilizing the spin structures are.

Effects that could suppress ordinary collinear magnetism and give way to finite-$Q$ order include geometric frustration in crystals with triangular motifs such as Gd$_{2}$PdSi$_{3}$ \cite{Okubo2012,Leonov2015,Kurumaji2019,Wang2021}, or frustration resulting from competition between interactions, such as inter-orbital exchange frustration \cite{Lin2016,Nomoto2020}. Alternatively, finite-$Q$ spin structures can be understood within the paradigm of itinerant magnetism as captured by the Kondo lattice model \cite{Inosov2009,Hayami2017,Ozawa2017,Wang2020,Yasui2020,Mitsumoto2021,Hayami2021a,Hayami2021b,Hayami2021c,Bouaziz2022,Dong2024}. Here, interactions between localized spins are mediated through coupling to the spins of itinerant electrons, and finite-$Q$ order is a consequence of Fermi surface nesting. The formation of non-collinear spin structures may then be expected to have a corresponding impact on the electronic structure. Indeed, electronic supermodulations and a possible nesting-driven instability have recently been observed to accompany non-collinear magnetism in GdRu$_{2}$Si$_{2}$ \cite{Yasui2020,Dong2025}, but a detailed relationship between charge and spin degrees-of-freedom remains to be established.

For GdRu$_{2}$Si$_{2}$ the magnetic phase diagram can be modeled by considering the Ruderman-Kittel-Kasuya-Yosida (RKKY) interaction (the lowest order term in the Hamiltonian of the Kondo lattice model), four-spin interactions (the 2$^{\mathrm{nd}}$ order term), and a Zeeman term \cite{Yasui2020}.
In contrast, for the sister compound GdRu$_{2}$Ge$_{2}$ it has recently been shown that despite the phase diagram's greater complexity, it can be successfully modeled using only the RKKY and Zeeman terms \cite{Yoshimochi2024}. The greater richness of the GdRu$_{2}$Ge$_{2}$ magnetic phase diagram is then thought to stem from a multiplicity of RKKY interactions occurring at different wavevectors, and the magnetic field-dependent changes in a competition between them. The electronic states that assist in stabilizing these phases may then be expected to exhibit a corresponding richness of multi-$Q$ superstructures.

\begin{figure*}
\centering
\includegraphics[scale=1]{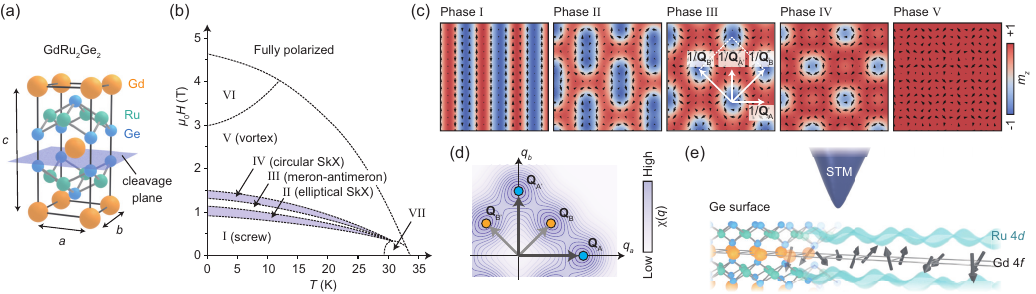}
\caption{\label{fig:1}
\textbf{Non-collinear magnetism in GdRu$_{2}$Ge$_{2}$.} (a) Depiction of the crystal structure with a cleavage plane indicated and, (b) a sketch of the magnetic phase diagram reported by Yoshimochi \textit{et al.} \cite{Yoshimochi2024}. (c) For magnetic phases I--V, the proposed spin textures on the Gd square lattice, inferred using resonant x-ray scattering and magnetization measurements \cite{Yoshimochi2024}. (d) Sketch of the bare magnetic susceptibility $\chi(q)$ showing maxima at $\mathbf{Q}_{\mathrm{A}(\mathrm{A}')}$ and $\mathbf{Q}_{\mathrm{B}(\mathrm{B}')}$. (e) Illustration of the rationale for this investigation: The layered structure of GdRu$_{2}$Ge$_{2}$ can be thought of as interleaved Gd magnetic layers and Ru-Ge conduction layers through which magnetic interactions are mediated. The itinerant electronic states in the Ru-Ge network are directly accessible to visualization using STM at the cleaved Ge surface.}
\end{figure*}

The structure of GdRu$_{2}$Ge$_{2}$ and a sketch of its magnetic phase diagram are shown in Figs. 1(a) and 1(b). The ground state, Phase~I, is proposed to exhibit a coplanar spin `screw' that propagates with wavevector $\mathbf{Q}_{\mathrm{A}}$, parallel to a lattice vector, as shown in Fig. 1(c). This is followed by a series of non-coplanar spin structures, Phases~II--V, each described by multiple interfering screw or screw-like modulations with wavevectors $\mathbf{Q}_{\mathrm{A}(\mathrm{A}')}$, $\mathbf{Q}_{\mathrm{B}(\mathrm{B}')}$ or $2\mathbf{Q}_{\mathrm{B}(\mathrm{B}')}$, where $\mathbf{Q}_{\mathrm{B}(\mathrm{B}')}$ are oriented diagonally with respect to the lattice vectors.
Phase~II and Phase~IV are skyrmion crystal phases, while Phase~III and Phase~V are meron-antimeron and magnetic vortex phases, respectively \cite{Yoshimochi2024}.
Approximations to these spin structures are plotted in Fig. 1(c).
Figure 1(d) depicts the bare susceptibility $\chi(q)$ where the local maxima at $\mathbf{Q}_{\mathrm{A}(\mathrm{A}')}$ and $\mathbf{Q}_{\mathrm{B}(\mathrm{B}')}$ are marked. 

Because the magnetic interactions are thought to be mediated by Ru 4$d$-derived bands residing within the Ru-Ge network \cite{Nomoto2020,Bouaziz2022,Rathnaweera2025}, they can be probed using scanning tunneling microscopy (STM) at Ge-terminated surfaces as depicted in Fig. 1(e).

\section*{Results}

\subsection*{Cleaved surfaces of GdRu$_{2}$Ge$_{2}$}

Topographic STM images of cleaved Ge- and Gd-terminated surfaces are shown in Figs. 2(a) and 2(b), respectively.
Figure 2(c) shows typical tunneling conductance spectra acquired in clean regions at each surface.
Aside from a sparse distribution of atomic-scale pits, the Gd surface is almost featureless in STM images, and is less interesting for the purposes of this work as the RKKY-mediating Ru bands are inaccessible. From this point onward we therefore focus entirely on the Ge surface. Due to the presence of cleavage debris, an extensive search for the largest clean, square field-of-view yielded only an $8 \times \SI{8}{nm^{2}}$ area, shown in Fig. 2(d).

\begin{figure*}
\centering
\includegraphics[scale=1]{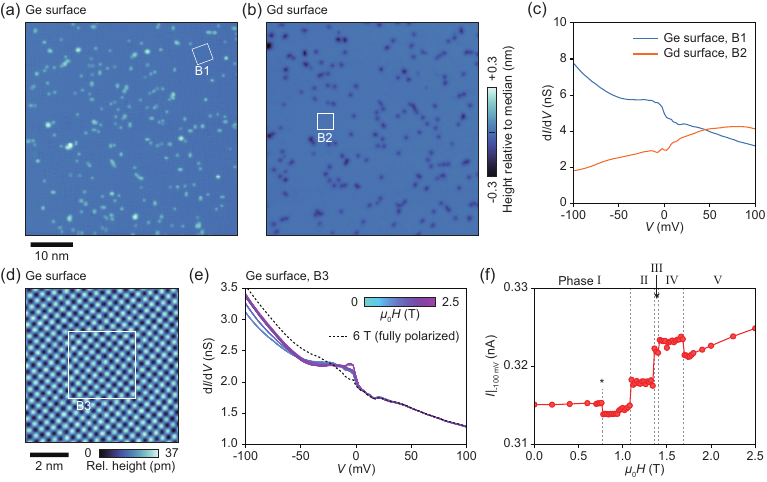}
\caption{\label{fig:2}
\textbf{STM at the cleaved surfaces of GdRu$_{2}$Ge$_{2}$.} (a) Topography at the Ge-terminated surface and, 	(b), at the Gd-terminated surface ($V_{\mathrm{set}} = \SI{100}{mV}$, $I_{\mathrm{set}} = \SI{200}{pA}$).
The Ge surface shows a sparse covering of atoms and clusters while the Gd surface shows a roughly corresponding density of atomic scale pits, and it is reasonable that the atoms and clusters on the Ge surface are Gd pulled from the opposing surface during cleavage. Both images are displayed with a shared color scale. (c) Tunneling conductance curves averaged over the boxed regions labeled as B1 and B2 in (a) and (b) respectively ($V_{\mathrm{set}} = \SI{100}{mV}$, $I_{\mathrm{set}} = \SI{200}{pA}$, $V_{\mathrm{mod}} = \SI{2.5}{mV}$, $\mu_{0}H = \SI{0}{T}$). (d) A higher resolution topographic image of the Ge surface ($V_{\mathrm{set}} = \SI{-100}{mV}$, $I_{\mathrm{set}} = \SI{200}{pA}$). The corrugations correspond to the Ge square net. (e) Magnetic field-dependence of the average conductance curve measured in the boxed region B3 in (d). (f) Magnetic field-dependence of tunneling current recorded at $V$ = $\SI{-100}{mV}$ after the tip height is stabilized at $V_{\mathrm{set}} = \SI{100}{mV}$ (for comparison with a similar quantity previously measured for GdRu$_{2}$Si$_{2}$ \cite{Yasui2020}). Subtle changes in this current correspond to meta-magnetic transitions between the phases described in Fig. 1. The change appearing within Phase~I, marked with an asterisk, is only partially reproducible and its origin is unknown.
}
\end{figure*}

The magnetic field-dependence of the tunneling spectrum acquired at the Ge surface is plotted in Fig. 2(e), showing only subtle variations. A quantity that emphasizes these variations, namely the tunneling current at opposite polarity to the setpoint, is shown in Fig. 2(f). Sharp jumps in this current coincide closely with the field strengths expected for the transitions between the magnetic phases described in Fig. 1 \cite{Yoshimochi2024}.

\begin{figure*}
\centering
\includegraphics[scale=1]{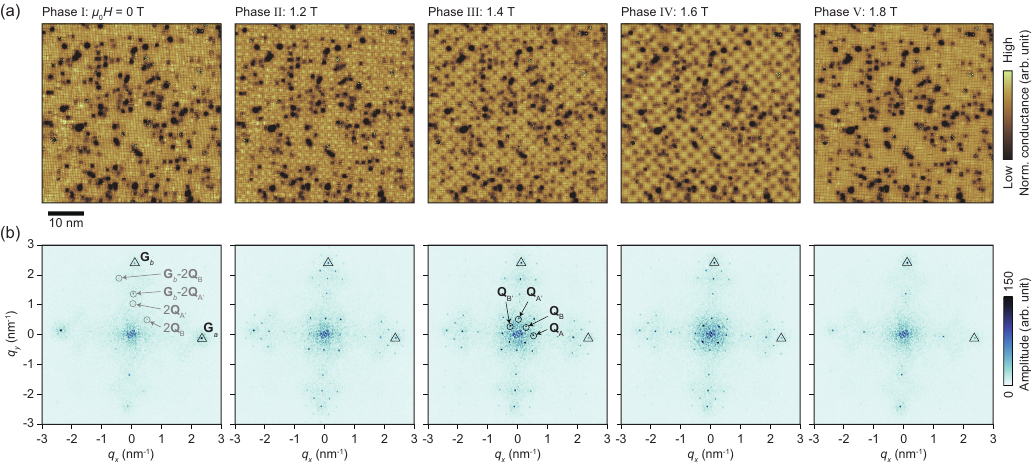}
\caption{\label{fig:3}
\textbf{Magnetic field-dependence of periodic modulations in the LDOS at the Ge surface.} (a) Normalized conductance images, $L(\mathbf{r}, V=\SI{-100}{mV})$, acquired for each of the magnetic phases I--V, at the indicated magnetic fields and in the same field-of-view as for Fig. 2(a) ($V_{\mathrm{set}} = \SI{-100}{mV}$, $I_{\mathrm{set}} = \SI{200}{pA}$, $V_{\mathrm{mod}} = \SI{5}{mV}$).
(b) Fast Fourier transforms, $\mathcal{F} \left[ L(\mathbf{r}, V=\SI{-100}{mV}) \right]$, obtained from the images in (a) after masking out the cleavage debris (see Supplementary Information Sec.~I). Bragg peaks are marked with black triangles. In the image for Phase~III we mark the prominent Fourier components associated with magnetic modulations $\mathbf{Q}_{\mathrm{A}(\mathrm{A}')}$ and $\mathbf{Q}_{\mathrm{B}(\mathrm{B}')}$ [recall Fig. 1(d)]. In the image for Phase~I we also mark Fourier components that do not correspond to the single wavevector expected to describe a spin screw. Their existence motivates a revision to the ground state magnetic structure, which we discuss below.
}
\end{figure*}

\subsection*{Imaging magnetic field-dependent LDOS textures}

As well as topography, STM can be used to image the local tunneling conductance which, after suitable normalization, is a fairly good proxy for the surface LDOS at energy $E = E_{\mathrm{F}} + eV$ (see Methods section for details). We denote this normalized tunneling conductance as $L(\mathbf{r}, V)$. In Fig. 3(a) we plot $L(\mathbf{r}, V=\SI{-100}{mV})$ in approximately the same field-of-view as for Fig. 2(a), and for applied fields that stabilize each magnetic phase. In each phase, a distinct periodic LDOS pattern appears. Fast Fourier transforms (FFTs), denoted $\mathcal{F} \left[ L(\mathbf{r}, V=\SI{-100}{mV}) \right]$, are shown in Fig. 3(b).
As well as the Bragg peaks, labeled $\mathbf{G}_{a}$ and $\mathbf{G}_{b}$, these show peaks at shorter wavevectors describing periodic superstructures in the LDOS. We see peaks associated with the magnetic modulation wavevectors $\mathbf{Q}_{\mathrm{A}(\mathrm{A}')}$ and $\mathbf{Q}_{\mathrm{B}(\mathrm{B}')}$ and their respective $2^{\mathrm{nd}}$ harmonics. We also see satellites of the Bragg peaks, such as $\mathbf{G}_{a}$-$\mathbf{Q}_{\mathrm{A}}$. In Phase~I, there is no peak that simply corresponds to a spin screw at $\mathbf{Q}_{\mathrm{A}}$. This by itself is not unexpected, but the appearance of wavevectors involving $\mathbf{Q}_{\mathrm{B}}$ is unexpected. It suggests that there is at least one magnetic modulation other than the spin screw, and this motivates a revision to the previously proposed ground state spin structure \cite{Yoshimochi2024}, which we implement for the model considered later in this work.

For each magnetic phase, the LDOS modulations and therefore the underlying magnetic modulations are coherent over lengthscales of several tens of nanometers. The measured ratios between the wavevectors $\mathbf{G}_{a(b)}$ and $\mathbf{Q}_{\mathrm{A}(\mathrm{A}')}$, estimated by fitting to the peaks in Fig. 3(b), indicate that the modulations deviate slightly from commensurability with the lattice (see Supplementary Information Sec.~II).

Figure 3 gives only a snapshot of the LDOS patterns seen in each magnetic phase. Next, in order to track changes in the patterns as a function of $H$ and resolve the magnetic phase transitions, we conduct further imaging of $L(\mathbf{r}, V)$ in the smaller field-of-view shown in Fig. 2(d).
Figure 4(a) shows $L(\mathbf{r}, V=\SI{-100}{mV})$ images extracted from a series acquired in increments of $\SI{20}{mT}$ throughout the range $ \mu_{0} H = 0.7$ to $\SI{1.8}{T}$. From Fig.~4(a) we see that although the patterns in Phases~II, III and IV have supercells of roughly the same size, their intra-supercell motifs are clearly distinct. 
An essential characteristic of multi-$Q$ patterns is that their $q$-space description must include not only the amplitudes of the constituent modulations [as in Fig.~3(b)], but also the modulations' respective phases, or more precisely the relations among their phases. Acquiring this information using an ordinary FFT runs into technical obstacles, so instead we use a non-uniform discrete Fourier transform (NUDFT, see Supplementary Information Sec.~III). In this way we precisely address the set of wavevectors that could in principle be physically relevant: The Bragg peaks $\mathbf{G}_{a(b)}$, the expected magnetic wavevectors $\mathbf{Q}_{\mathrm{A}(\mathrm{A}')}$ and $\mathbf{Q}_{\mathrm{B}(\mathrm{B}')}$, their $2^{\mathrm{nd}}$ harmonics, and the satellites and satellite $2^{\mathrm{nd}}$ harmonics stemming from $\mathbf{G}_{a(b)}$. In total there are 42 unique wavevectors to be addressed, referred to collectively as $\mathcal{Q}$. We use a NUDFT to acquire amplitude and phase information at those wavevectors, and finally remove unimportant phase offsets for ease of comparison and interpretation. This also references all phases against a corner of the magnetic unit-cell and, simultaneously, the location of a Ge ion somewhere, and therefore also the Gd ion beneath it. (This procedure is described in Supplementary Information Sec.~III.) The result is a selective Fourier spectrum for a given image, denoted $\tilde{\mathcal{F}}_{\mathcal{Q}}[L(\mathbf{r})]$, examples of which are shown in Fig.~4(b).

\begin{figure*}
\centering
\includegraphics[scale=1]{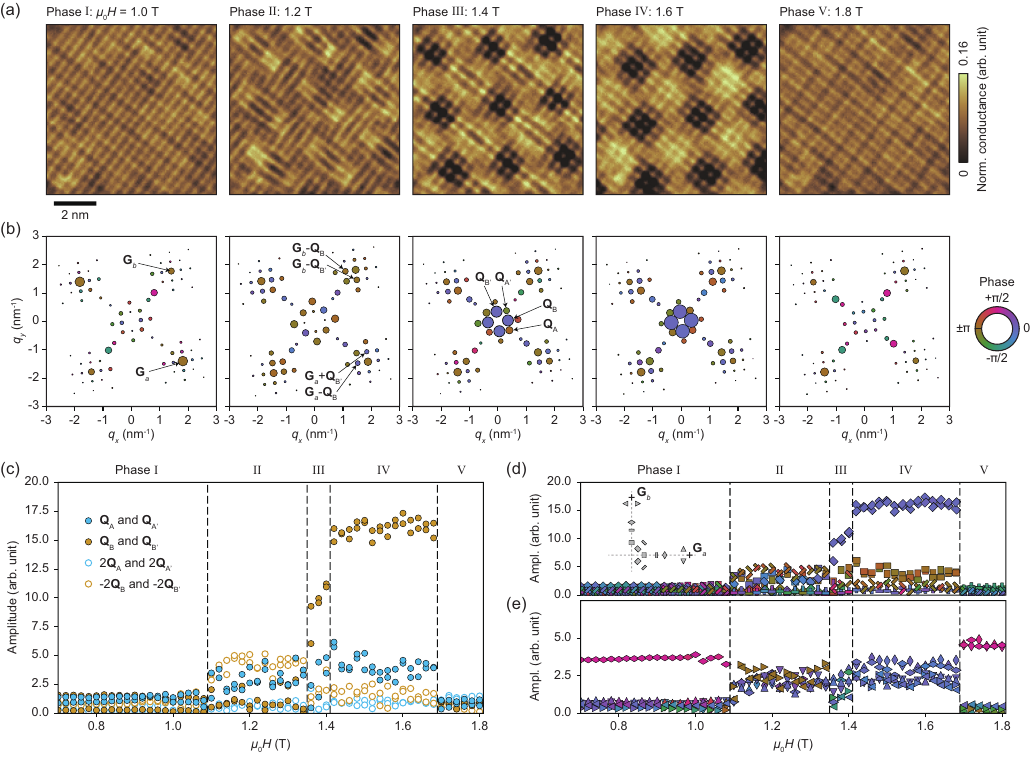}
\caption{\label{fig:4}
\textbf{Fourier spectra of measured LDOS motifs and their magnetic field-dependence.} (a) High-resolution $L(\mathbf{r})$ images for each magnetic phase in the same field-of-view as for Fig. 2(d) ($V_{\mathrm{set}} = \SI{-100}{mV}$, $I_{\mathrm{set}} = \SI{200}{pA}$, $V_{\mathrm{mod}} = \SI{5}{mV}$). (b) Sets of selective Fourier spectra, $\tilde{\mathcal{F}}_{\mathcal{Q}}[L(\mathbf{r})]$, obtained from the images in (a) using a NUDFT. The area of each marker in the plots corresponds to the amplitude of the given complex Fourier coefficient. (The area of the smallest markers reflects the $q$-independent noise floor.) The marker color corresponds to the coefficient's phase, as shown in the colour wheel on the right hand side. For Phase~II, the satellite components ($\mathbf{G}_{a}\!-\!\mathbf{Q}_{B}$, etc.) that encode the peculiar basketweave pattern are marked.
(c) The amplitudes of the $\mathbf{Q}_{\mathrm{A}(\mathrm{A}')}$ and $\mathbf{Q}_{\mathrm{B}(\mathrm{B}')}$ components, and of their $2^{\mathrm{nd}}$ harmonics. (d) Amplitudes and phases for the same subset of Fourier components and, (e), amplitudes and phases for components other than those that straightforwardly correspond to the expected magnetic modulations, but which nonetheless capture the distinct character of each observed LDOS pattern. The marker colour corresponds to phase as indicated by the colour wheel in (b).
As well as magnetic field-dependence, the energy-dependence of $L(\mathbf{r})$ and of $\tilde{\mathcal{F}}_{\mathcal{Q}}[L(\mathbf{r})]$ are presented in Supplementary Information Sec.~IV.
}
\end{figure*}

In Fig. 4(a), Phase~II exhibits a pattern that we refer to as  the `basketweave' pattern. It is composed of a pair of bond-order-like stripe components which are themselves spatially modulated in anti-phase with each other. The Fourier components which capture this behavior are satellites ($\mathbf{G}_{a}\!-\!\mathbf{Q}_{\mathrm{B}}$, etc.) marked in the NUDFT plot for Phase~II, and discussed more fully in Supplementary Information Sec.~V. This texture is especially noteworthy as its detailed bond-order-like features allow us to establish the $r$-space microscopic units that build up all the $L(\mathbf{r})$ images in Fig. 4(a), as we will show below.

Having measured both the LDOS modulations' amplitudes and phases throughout the range $ \mu_{0} H = 0.7$ to $\SI{1.8}{T}$, we can observe how they vary with magnetic field. First, Fig. 4(c) simply shows the amplitudes or Fourier components that correspond to the principle magnetic modulations at $\mathbf{Q}_{\mathrm{A}(\mathrm{A}')}$ and $\mathbf{Q}_{\mathrm{B}(\mathrm{B}')}$, and their $2^{\mathrm{nd}}$ harmonics. Na\"{i}vely, the amplitudes at $\mathbf{Q}_{\mathrm{A}(\mathrm{A}')}$ and $\mathbf{Q}_{\mathrm{B}(\mathrm{B}')}$ may reflect the electronic susceptibility, and the changing competition between RKKY interactions, at those wavevectors.

In Figs. 4(d) and 4(e) we show the magnetic field dependence of a more complete set of amplitudes and phases characterizing the motifs in Fig. 4(a). These are fairly consistent within each magnetic phase, and change abruptly at the magnetic phase transitions. As expected, the transitions are seen to be $1^{\mathrm{st}}$ order in nature: Upon repetition of a measurement series of the kind shown in Fig. 4, the transitions occur at different magnetic field values in each sweep (see Supplementary Information Sec.~VI).

As the phases shown in Figs. 4(b) and 4(d) are referenced against a spatial location corresponding to a particular Gd atom, they contain information about the intra-supercell structure down to its registry with the atomic lattice and bond-centered sublattices. This provides concrete markers to assess how well a given numerical scheme models the observed patterns. We introduce a simple example of such a scheme, and its resulting $r$-space maps and Fourier spectra, in the remainder of this article.

\begin{figure}
\centering
\includegraphics[scale=1]{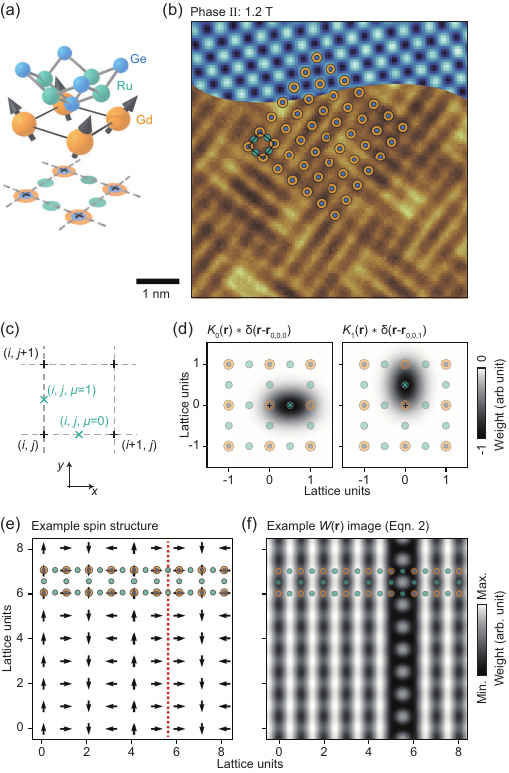}
\caption{\label{fig:5}
\textbf{Scheme for modeling \textit{L} images given the proposed spin structures.} (a) Illustration of the near-surface atomic structure for the Ge termination, and the planar projection of the atom locations most relevant here. (b) A comparison of simultaneously measured STM topography and $L(\mathbf{r})$ images for Phase~II. The intensity of $L$ exhibits `trenches' along the Gd-Ru-Gd bond chains. 
(c) The notation scheme for specifying Gd sites and the pair of inequivalent bond centers associated with each. The index $\mu \in \{ 0,1 \}$ specifies the bond-centered site relative to the site $(i,j)$, so that $(i,j,\mu)$ specifies a bond. This also specifies the location of one of the Ru ions (projected to the Gd plane), of which there are two per unit-cell. 
(d) Kernel functions designed to capture the observed intra-unit cell distribution. Each is a $90^{\circ}$-rotated versions of the other, to account for the perpendicular bond orientations. Here each kernel $K_{\mu}(\mathbf{r})$ is convolved with a delta function $\delta(\mathbf{r}\!-\!\mathbf{r}_{i,j,\mu})$ representing the relevant bond center. Note that they are negative-valued and so describe a \textit{negative} NCMR effect. (e) A fictitious spin structure in which the nearest-neighbor dot products are generally $0$ for one bond sublattice and $1$ for the other. A phase-slip defect is included (red dotted line), creating a ladder of fully aligned spins. (f) The weight image $W(\mathbf{r})$ calculated using Eqns. 1 and 2 for the spin structure in (e).
}
\end{figure}

\subsection*{Modeling magnetic field-dependent LDOS textures}

Previously, several effects have been considered in order to understand the impact of non-collinear magnetism on STM conductance images.
In the tunneling magneto-resistance (TMR) effect (the operative effect in spin-polarized STM measurements \cite{Heinze2011,Spethmann2024}), varying local spin orientations impact the junction conductance \textit{via} the varying tunneling probability \cite{Julliere1975}. In this work, the use of a non-magnetic STM tip rules out the TMR effect, but even in this case other mechanisms have been reported which create contrast in conductance images, such as the tunneling anisotropic magneto-resistance (TAMR) \cite{Gould2004} and non-collinear magneto-resistance (NCMR) effects \cite{Hanneken2015,Kubetzka2017}. These phenomena are distinct from TMR because they impact the conductance not through varying tunneling probabilities, but rather through the variation of the spin-integrated DOS or LDOS intrinsic to a junction's magnetic electrode (in this case the sample surface). The TAMR effect is ruled out by our observations of the ground state: It would be expected to cause an LDOS modulation with wavevector $\mathbf{Q}_{\mathrm{A}}$, which is not observed.

Another effect, considered by Yasui \textit{et al.} to understand LDOS modulations in GdRu$_{2}$Si$_{2}$, is the formation of a charge density wave in an electron band due to interaction with localized spins \textit{via} a Kondo-type exchange coupling \cite{Yasui2020,Hayami2021}. There, the charge density was resolved only at the sites of the localized spins, i.e. only on the Gd lattice. Although this approach successfully models some qualitative features of the observed LDOS images, it is somewhat of a `black box': the microscopic relationship between charge and spin remains unclear.
Furthermore, a more microscopically detailed treatment is now required, most obviously for the basketweave configuration seen in Phase II. Here the nominal $C_{4v}$ lattice symmetry is broken both broadly within the motif of the magnetic unit-cell and also at the level of the crystal unit-cell, as emphasized in Figs. 5(a) and 5(b). Therefore, not only the Gd site-centred local electron density, but also the bond-centred and hollow-centred local densities must be addressed.

\begin{figure*}
\centering
\includegraphics[scale=1]{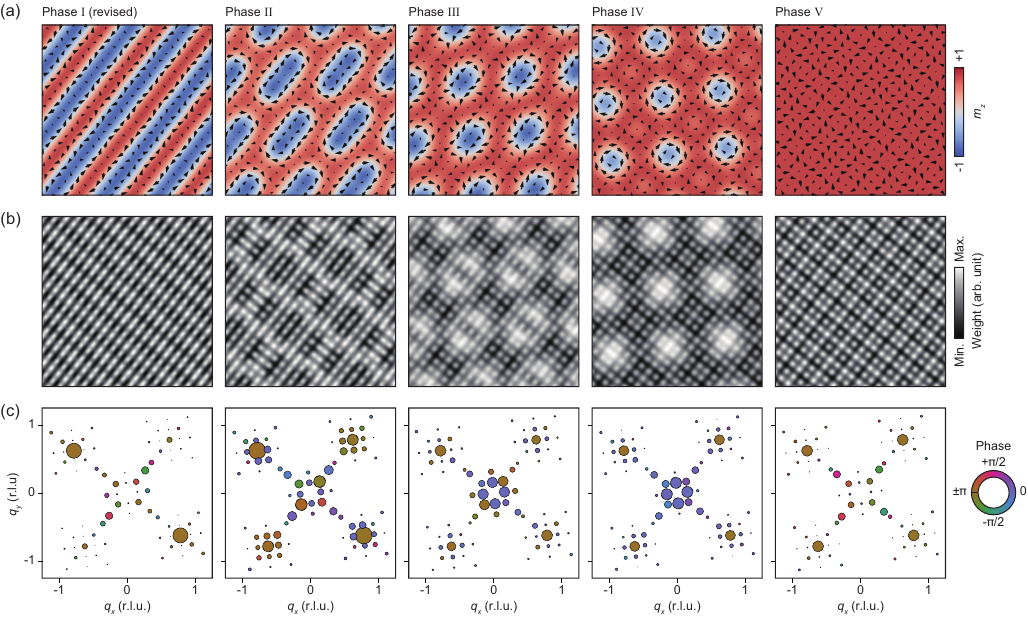}
\caption{\label{fig:6}
\textbf{Modeled \textit{W}(\textbf{r}) maps and their Fourier spectra.} (a) Magnetic structures used in the modeling. Note that the structure for Phase~I has been modified from that reported by Yoshimochi \textit{et al.} \cite{Yoshimochi2024}, as described in detail in Supplemental Information Sec.~VII. It now includes an additional magnetic modulation, an in-plane spin sinusoid propagating perpendicular to the original spin screw, and resembles that recently suggested for GdRu$_{2}$Si$_{2}$ \cite{Khanh2022,Wood2023}. It also exhibits similar topological charge stripes  -- see Supplementary Information Sec.~VIII. (b) Modeled bond weight images $W(\mathbf{r})$ obtained using Eqns. 1 and 2. (c) Fourier spectra $\tilde{\mathcal{F}}_{\mathcal{Q}}[W(\mathbf{r})]$ obtained using the same processes as for the measured $L(\mathbf{r})$ data as shown in Fig. 4.
}
\end{figure*}

Having ruled out the TMR and TAMR effects, we adopt the basic premise of the NCMR effect: The LDOS depends in some way on the nearest-neighbor spin alignments as captured by the spin-spin dot products, for example $\mathbf{S}_{i,j} \cdot \mathbf{S}_{i+1,j}$ where $i$ and $j$ index sites on the Gd square lattice. If $\mu \in \{ 0,1 \}$ indexes the two bonds associated with the site $(i,j)$ as shown in Fig. 5(c), then the spin-spin product associated with bond $(i,j,\mu)$ is
\begin{equation}
P_{i,j,\mu} = (1 - \mu) (\mathbf{S}_{i,j} \cdot \mathbf{S}_{i+1,j}) + \mu(\mathbf{S}_{i,j} \cdot \mathbf{S}_{i,j+1}).
\end{equation}
We expand upon this lattice-based information to obtain continuous-domain information, comparable to a high-resolution STM image, using Fourier convolution. We first place an array of delta functions in $r$-space, $\delta(\mathbf{r}\!-\!\mathbf{r}_{i,j,\mu})$, representing sites of the two bond-centred lattices. These are then convolved with suitable bond-dependent kernel functions $K_{\mu}(\mathbf{r})$. Finally, the contribution of any given bond to a resulting `weight' image $W(\mathbf{r})$ is set according to the product $P_{i,j,\mu}$, so that
\begin{equation}
W(\mathbf{r}) = \sum_{i,j,\mu} \Big[ \left[ P_{i,j,\mu} K_{\mu}(\mathbf{r}) \right] * \delta(\mathbf{r}-\mathbf{r}_{i,j,\mu}) \Big],
\end{equation}
where `$*$' denotes Fourier convolution.

The choice of the kernel functions $K_{\mu}(\mathbf{r})$ is informed by a close inspection of the basketweave pattern. In Fig. 5(b) we show the registry between the topographic corrugations, in which the peaks correspond to Ge sites and the Gd sites below them, and $L(\mathbf{r}, V=\SI{-100}{mV})$. The maxima of $L$ tend to appear at the hollow centers of the Gd lattice, and the minima tend to appear as elongated `trenches' connecting the Gd sites along the Gd-Ru-Gd bond lines, creating local striped patterns in relief. From this we infer that the bond-centred kernels should be i) negative-valued and ii) elongated along the bond axis.
The chosen kernel functions are shown in Fig. 5(d). For the sake of example, a simple fictitious spin structure and the resulting $W(\mathbf{r})$ image calculated using Eqns. 1 and 2, are shown in Figs. 5(e) and 5(f) respectively.

Figure 6 shows the results of applying the above treatment to the spin structures of Phases~I--V of GdRu$_{2}$Ge$_{2}$. Note that here we consider a revised spin structure for Phase~I, which includes an additional modulation propagating perpendicular to the spin screw. This is motivated by the observation of unexpected modulations highlighted in Fig. 3(b). (Further details are given in the Supplementary Information Sec.~VII.) These spin structures are shown in Fig. 6(a), suitably rotated and shifted for comparison with the observations shown in Fig. 4. The calculated $W(\mathbf{r})$ images are shown in Fig. 6(b), and their Fourier spectra $\tilde{\mathcal{F}}_{\mathcal{Q}}[W(\mathbf{r})]$ are shown in Fig. 6(c). 

A direct comparison can be drawn between the calculated images in Figs. 6(b) and 6(c), and the corresponding measurement results shown in Figs. 4(a) and 4(b). For each phase, the broad qualitative features are reproduced reasonably well. The bond-order-like stripes with a low-$q$ super-modulation in Phase~I are well represented, as are the essential features of the basketweave pattern in Phase~II -- a periodic arrangement of patches of bond-order-like stripes that alternate between two perpendicular orientations. The checkerboard-like pattern in Phase~IV is well represented, as well as the pattern in Phase~III which is similar, but has a slight suppression of intensity at the centers of the bright regions. Finally, the pattern in Phase~V, which resembles a sum of two copies the pattern in Phase~I, orientated perpendicular to each other, is also successfully modeled. The phases of Fourier coefficients in the spectra $\tilde{\mathcal{F}}_{\mathcal{Q}}[W(\mathbf{r})]$ and $\tilde{\mathcal{F}}_{\mathcal{Q}}[L(\mathbf{r})]$ also match reasonably well. In particular, the configuration of phases of the Fourier coefficients at the satellites $\mathbf{G}_{a}\!-\!\mathbf{Q}_{\mathrm{B}}$, etc., which encode the intricate interplay of constructive and destructive interference resulting in the basketweave pattern, are accurately represented. (See Supplementary Information Sec.~V for more detail.) A notable discrepancy is that the model generates Fourier components at wavevectors such as $\mathbf{Q}_{\mathrm{A}}\!+\!\mathbf{Q}_{\mathrm{B}}$ which do not appear in the Fourier transforms of $L(\mathbf{r})$ images shown in Fig. 3. This is shown and discussed in the Supplementary Information, Sec.~IX.

\section*{Discussion}

The observations presented above show that the series of multi-$Q$ magnetic phases in GdRu$_{2}$Ge$_{2}$ are accompanied by corresponding multi-$Q$ patterns in the LDOS at an energy near $E_{\mathrm{F}}$. This indicates distinct reconstructions of the electronic bands with the onset of each magnetic phase. The tunneling conductance curves in Fig. 2(e) show that all magnetic Phases I--V exhibit some transfer of spectral weight in comparison with the curve for the fully polarized state measured at $\mu_{0}H = \SI{6}{T}$, also consistent with band reconstructions. These reconstructions and the accompanying shrinking of the Brillouin zone can be expected to impact the interactions that stabilize the magnetism in the first place -- a complicated chicken-and-egg scenario. Although the observations in this work support the central importance of itinerant electrons in the formation of the magnetic structures, it is noteworthy that a double-$Q$ ground state does require the presence of mechanisms beyond RKKY, such as four-spin interactions \cite{Hayami2017,Khanh2022}.

Alongside the $q$-space description of the LDOS patterns, the analysis above provides an understanding of their microscopic composition in $r$-space. They are well-modeled using anisotropic kernels at bond-centred sites of the Gd lattice or, equivalently, the Ru sites. This finding aligns well with the understanding that interactions between Gd spins are mediated by electronic states that are mostly derived from Ru 4$d$ orbitals. Indeed, the forms of the kernel functions adopted above [Fig. 5(d)] resemble the planar projections that might be expected for Ru $d_{xz}$ and $d_{yz}$ orbitals. The contribution of these Ru-centred objects to the LDOS images is shown to depend primarily on the alignment of the Gd spins between which the Ru site is located. This establishes a concrete microscopic relationship between the spin and charge degrees of freedom involved in the emergence of itinerant topological magnetism.

\section*{Methods}

Single crystal samples were synthesized as described previously \cite{Yoshimochi2024}. They were loaded into an ultra-high vacuum chamber ($P \sim \SI{e-10}{Torr}$) before cleaving at about $\SI{77}{K}$, after which they were inserted into a modified Unisoku USM1300 low-temperature STM system \cite{Hanaguri2006}. STM measurements were performed using an electro-chemically etched tungsten tip that was characterized and conditioned using field ion microscopy followed by repeated mild indentation at a clean Cu(111) surface. All measurements were performed at $T = \SI{1.5}{K}$. Tunneling conductance, $\frac{\mathrm{d}I}{\mathrm{d}V}(V)$,  was measured using the lock-in technique with bias modulation of frequency $f_{\textrm{mod}} = \SI{617.3}{Hz}$ and an amplitude $V_{\mathrm{mod}}$ specified in the caption describing each measurement.

To obtain images of the approximate LDOS, as shown in Figs. 3 and 4, we first acquire tunneling conductance $\frac{\mathrm{d}I}{\mathrm{d}V}(\mathbf{r},V)$ and, to mitigate artifacts of STM tip height variations, compute the normalized conductance $L(\mathbf{r}, V) = [\frac{\mathrm{d}I}{\mathrm{d}V}(\mathbf{r},V)]/[I(\mathbf{r}, V)/V]$ \cite{Kohsaka2007,Macdonald2016}.

STM topography maps, $L(\mathbf{r}, V)$ maps, $\mathcal{F} \left[ L(\mathbf{r}, V) \right]$ images, NUDFT results and $W(r)$ images are plotted using perceptually uniform colormaps \cite{Thyng2016}.

\section*{Acknowledgements}
We are grateful to  T. Machida and M. Naritsuka for assistance. This work was supported by JSPS KAKENHI Grants No. JP21H04990, JP22KJ1061, JP22H04965, JP24K00579, JP24H02235, JP25K00956, and 25H00611, a JSPS Grant-in-Aid for Scientific Research on Innovative Areas `Quantum Liquid Crystals' (KAKENHI Grant No. JP19H05824), JST CREST Grant No. JPMJCR23O4, JST PRESTO Grant No. JPMJPR20B4, and by the Asahi Glass Foundation, the Murata Science Foundation, and the Noguchi Institute.

\section*{Data availability}
The data that support the findings presented here are available from the corresponding authors upon reasonable request.


\begin{thebibliography}{99}


\bibitem{Dzyaloshinskii1958}
I. Dzyaloshinskii,
\textit{A thermodynamic theory of ``weak'' ferromagnetism of antiferromagnetics.}
J. Phys. Chem. Solids
\textbf{4},
241--255
(1958).
\url{https://doi.org/10.1016/0022-3697(58)90076-3}


\bibitem{Moriya1960}
T. Moriya,
\textit{Anisotropic Superexchange Interaction and Weak Ferromagnetism.}
Phys. Rev.
\textbf{120},
91--98
(1960).
\url{https://doi.org/10.1103/PhysRev.120.91}


\bibitem{Muhlbauer2009}
S. M\"{u}hlbauer, B. Binz, F. Jonietz, C. Pfleiderer, A. Rosch, A. Neubauer, R. Georgii, and P. B\"{o}ni,
\textit{Skyrmion Lattice in a Chiral Magnet.}
Science
\textbf{323},
915--919
(2009).
\url{https://doi.org/10.1126/science.1166767}


\bibitem{Yu2010}
X. Z. Yu, Y. Onose, N. Kanazawa, J. H. Park, J. H. Han, Y. Matsui, N. Nagaosa, and Y. Tokura,
\textit{Real-space observation of a two-dimensional skyrmion crystal.}
Nature
\textbf{465},
901--904
(2010).
\url{https://doi.org/10.1038/nature09124}


\bibitem{Heinze2011}
S. Heinze, K. von Bergmann, M. Menzel, J. Brede, A Kubetzka, R. Wiesendanger, G. Bihlmayer, and S. Bl\"{u}gel,
\textit{Spontaneous atomic-scale magnetic skyrmion lattice in two dimensions.}
Nature Physics
\textbf{7},
713--718
(2011).
\url{https://doi.org/10.1038/nphys2045}


\bibitem{Khanh2015}
N. D. Khanh, T. Nakajima, X. Yu, S. Gao, K. Shibata, M. Hirschberger, Y. Yamasaki, H. Sagayama, H. Nakao, L. Peng, K. Nakajima, R. Takagi, T. Arima, Y. Tokura, and S. Seki,
\textit{Nanometric square skyrmion lattice in a centrosymmetric tetragonal magnet.}
Nature Nanotechnology
\textbf{15},
444--449
(2015).
\url{https://doi.org/10.1038/s41565-020-0684-7}


\bibitem{Kurumaji2019}
T. Kurumaji, T. Nakajima, M. Hirschberger, A. Kikkawa, Y. Yamasaki, H. Sagayama, H. Nakao, Y. Taguchi, T.-h. Arima, and Y. Tokura,
\textit{Skyrmion lattice with a giant topological Hall effect in a frustrated triangular-lattice magnet.}
Science
\textbf{365},
914--918
(2019).
\url{https://doi.org/10.1126/science.aau0968}


\bibitem{Hirschberger2019}
M. Hirschberger, T. Nakajima, S. Gao, L. Peng, A. Kikkawa, T. Kurumaji, M. Kriener, Y. Yamasaki, H. Sagayama, H. Nakao, K. Ohishi, K. Kakurai, Y. Taguchi, X. Yu, T.-h. Arima, and Y. Tokura,
\textit{Skyrmion phase and competing magnetic orders on a breathing kagom\'{e} lattice.}
Nature Communications
\textbf{10},
5831
(2019).
\url{https://doi.org/10.1038/s41467-019-13675-4}


\bibitem{Ishiwata2020}
S. Ishiwata, T. Nakajima, J.-H. Kim, D. S. Inosov, N. Kanazawa, J. S. White, J. L. Gavilano, R. Georgii, K. M. Seemann, G. Brandt, P. Manuel, D. D. Khalyavin, S. Seki, Y. Tokunga, M. Kinoshita, Y. W. Long, Y. Kaneko, Y. Taguchi, T. Arima, B. Keimer, and Y. Tokura,
\textit{Emergent topological spin structures in the centrosymmetric cubic perovskite SrFeO$_{3}$.}
Phys. Rev. B
\textbf{101},
134406
(2020).
\url{https://doi.org/10.1103/PhysRevB.101.134406}


\bibitem{Gao2020}
S. Gao, H. D. Rosales, F. A. G. Albarrac\'{i}n, V. Tsurkan, G. Kaur, T. Fennell, P. Steffens, M. Boehm, P. \v{C}erm\'{a}k, A. Schneidewind, E. Ressouche, D. C. Cabra, C. R\"{u}egg, and O. Zaharko,
\textit{Fractional antiferromagnetic skyrmion lattice induced by anisotropic couplings.}
Nature
\textbf{586},
37--41
(2020).
\url{https://doi.org/10.1038/s41586-020-2716-8}


\bibitem{Takagi2022}
R. Takagi, N. Matsuyama, V. Ukleev, L. Yu, J. S. White, S. Francoual, J. R. L. Mardegan, S. Hayami, H. Saito, K. Kaneko, K. Ohishi, Y. \-{O}nuki, T.-h. Arima, Y. Tokura, T. Nakajima, and S. Seki,
\textit{Square and rhombic lattices of magnetic skyrmions in a centrosymmetric binary compound.}
Nature Communications
\textbf{13},
1472
(2022).
\url{https://doi.org/10.1038/s41467-022-29131-9}


\bibitem{Khanh2022}
N. D. Khanh, T. Nakajima, S. Hayami, S. Gao, Y. Yamasaki,
H. Sagayama, H. Nakao, R. Takagi, Y. Motome, Y. Tokura, T.-h. Arima, and S. Seki,
\textit{Zoology of multiple-$Q$ spin textures in a centrosymmetric tetragonal magnet with itinerant electrons.}
Adv. Sci.
\textbf{9},
2105452
(2022).
\url{https://doi.org/10.1002/advs.202105452}


\bibitem{Okubo2012}
T. Okubo, S. Chung, and H. Kawamura,
\textit{Multiple-q States and the Skyrmion Lattice of the Triangular-Lattice Heisenberg Antiferromagnet under Magnetic Fields.}
Phys. Rev. Lett.
\textbf{108},
017206
(2012).
\url{https://doi.org/10.1103/PhysRevLett.108.017206}


\bibitem{Leonov2015}
A. O. Leonov and M. Mostovoy,
\textit{Multiply periodic states and isolated skyrmions in an anisotropic frustrated magnet.}
Nature Communications
\textbf{6},
8275
(2015).
\url{https://doi.org/10.1038/ncomms9275}


\bibitem{Wang2021}
Z. Wang, Y. Su, S.-Z. Lin, and C. D. Batista,
\textit{Meron, skyrmion, and vortex crystals in centrosymmetric tetragonal magnets.}
Phys. Rev. B
\textbf{103},
104408
(2021).
\url{https://doi.org/10.1103/PhysRevB.103.104408}


\bibitem{Lin2016}
S.-Z. Lin and S. Hayami, 
\textit{Ginzburg-Landau theory for skyrmions in inversion-symmetric magnets with competing interactions.}
Phys. Rev. B
\textbf{93},
064430
(2016).
\url{https://doi.org/10.1103/PhysRevB.93.064430}


\bibitem{Nomoto2020}
T. Nomoto, T. Koretsune, and R. Arita,
\textit{Formation Mechanism of the Helical $Q$ Structure in Gd-Based Skyrmion Materials.}
Phys. Rev. Lett.
\textbf{125},
117204
(2020).
\url{https://doi.org/10.1103/PhysRevLett.125.117204}


\bibitem{Inosov2009}
D. S. Inosov, D. V. Evtushinsky, A. Koitzch, V. B. Zabolotnyy, S. V. Borisenko, A. A. Kordyuk, M. Frontzek, M. Loewenhaupt, W. L\"{o}ser, I. Mazilu, H. Bitterlich, G. Behr, J.-U. Hoffmann, R. Follath, and B. B\"{u}chner,
\textit{Electronic Structure and Nesting-Driven Enhancement of the RKKY Interaction at the Magnetic Ordering Propagation Vector in Gd$_{2}$PdSi$_{3}$ and Tb$_{2}$PdSi$_{3}$.}
Phys. Rev. Lett.
\textbf{102},
046401
(2009).
\url{https://doi.org/10.1103/PhysRevLett.102.046401}


\bibitem{Hayami2017}
S. Hayami, R. Ozawa, and Y. Motome,
\textit{Effective bilinear-biquadratic model for noncoplanar ordering in itinerant magnets.}
Phys. Rev. B
\textbf{95},
224424
(2017).
\url{https://doi.org/10.1103/PhysRevB.95.224424}


\bibitem{Ozawa2017}
R. Ozawa, S. Hayami, and Y. Motome,
\textit{Zero-Field Skyrmions with a High Topological Number in Itinerant Magnets.}
Phys. Rev. Lett.
\textbf{118},
147205
(2017).
\url{https://doi.org/10.1103/PhysRevLett.118.147205}


\bibitem{Wang2020}
Z. Wang, Y. Su, S.-Z. Lin, and C. D. Batista,
\textit{Skyrmion Crystal from RKKY Interaction Mediated by 2D Electron Gas.}
Phys. Rev. Lett.
\textbf{124},
207201
(2020).
\url{https://doi.org/10.1103/PhysRevLett.124.207201}


\bibitem{Yasui2020}
Y. Yasui, C. J. Butler, N. D. Khanh, S. Hayami, T. Nomoto, T. Hanaguri, Y. Motome, R. Arita, T. Arima, Y. Tokura, and S. Seki,
\textit{Imaging the coupling between itinerant electrons and localised moments in the centrosymmetric skyrmion magnet GdRu$_{2}$Si$_{2}$.}
Nature Communications
\textbf{11},
5925
(2020).
\url{https://doi.org/10.1038/s41467-020-19751-4}


\bibitem{Mitsumoto2021}
K. Mitsumoto and H. Kawamura,
\textit{Replica symmetry breaking in the RKKY skyrmion-crystal system.}
Phys. Rev. B
\textbf{104},
184432
(2021).
\url{https://doi.org/10.1103/PhysRevB.104.184432}


\bibitem{Hayami2021a}
S. Hayami and Y. Motome,
\textit{Square skyrmion crystal in centrosymmetric itinerant magnets}
Phys. Rev. B
\textbf{103},
024439
(2021).
\url{https://doi.org/10.1103/PhysRevB.103.024439}


\bibitem{Hayami2021b}
S. Hayami, T. Okubo, and Y. Motome,
\textit{Phase shift in skyrmion crystals.}
Nature Communications
\textbf{12},
6927
(2021).
\url{https://doi.org/10.1038/s41467-021-27083-0}


\bibitem{Hayami2021c}
S. Hayami and Y. Motome,
\textit{Topological spin crystals by itinerant frustration.}
J. Phys.: Condens. Matter
\textbf{33},
443001
(2021).
\url{https://doi.org/10.1088/1361-648X/ac1a30}


\bibitem{Bouaziz2022}
J. Bouaziz, E. Mendive-Tapla, S. Bl\"{u}gel, and J. B. Staunton,
\textit{Fermi-Surface Origin of Skyrmion Lattices in Centrosymmetric Rare-Earth Intermetallics.}
Phys. Rev. Lett.
\textbf{128},
157206
(2022).
\url{https://doi.org/10.1103/PhysRevLett.128.157206}


\bibitem{Dong2024}
Y. Dong, Y. Arai, K. Kuroda, M. Ochi, N. Tanaka, Y. Wan, M. D. Watson, T. K. Kim, C. Cacho, M. Hashimoto, D. Lu, Y. Aoji, T. D. Matsuda, and T. Kondo,
\textit{Fermi Surface Nesting Driving the RKKY Interaction in the Centrosymmetric Skyrmion Magnet Gd$_{2}$PdSi$_{3}$.}
Phys. Rev. Lett.
\textbf{133},
016401
(2024).
\url{https://doi.org/10.1103/PhysRevLett.133.016401}


\bibitem{Dong2025}
Y. Dong, Y. Kinoshita, M. Ochi, R. Nakachi, R. Higashinaka, S. Hayami, Y. Wan, Y. Arai, S. Huh, M. Hashimoto, D. Lu, M. Tokunaga, Y. Aoji, T. D. Matsuda, and T. Kondo,
\textit{Pseudogap and Fermi arc induced by Fermi surface nesting in a centrosymmetric skyrmion magnet.}
Science
\textbf{388},
624--630
(2025).
\url{https://doi.org/10.1126/science.adj7710}


\bibitem{Yoshimochi2024}
H. Yoshimochi, R. Takagi, J. Ju, N. D. Khanh, H. Saito, H. Sagayama, H. Nakao, S. Itoh, Y. Tokura, T. Arima, S. Hayami, T. Nakajima, and S. Seki,
\textit{Multistep topological transitions among meron and skyrmion crystals in a centrosymmetric magnet.}
Nature Physics
\textbf{20},
1001--1008
(2024).
\url{https://doi.org/10.1038/s41567-024-02445-9}


\bibitem{Rathnaweera2025}
D. N. Rathnaweera, X. Huai, K. R. Kumar, Y. Wang, J. Schlesinger, C. J. Bartel, S. Tewari, M. J. Winiarski, R. Dronskowski, and T. T. Tran,
\textit{Antibonding and electronic instabilities in GdRu$_{2}$X$_{2}$ (X = Si, Ge, and Sn): a new pathway toward developing centrosymmetric skyrmion materials.}
Journal of Materials Chemistry C
(2026).
\url{https://doi.org/10.1039/d5tc02333e}


\bibitem{Spethmann2024}
J. Spethmann, N. D. Khanh, H. Yoshimochi, R. Takagi, S. Hayami, Y. Motome, R. Wiesendanger, S. Seki, and K. von Bergmann,
\textit{SP-STM study of the multi-Q phases in GdRu$_{2}$Si$_{2}$.}
Phys. Rev. Mater.
\textbf{8},
064404
(2024).
\url{https://doi.org/10.1103/PhysRevMaterials.8.064404}


\bibitem{Julliere1975}
M. Julli\`{e}re,
\textit{Tunneling between ferromagnetic films.}
Phys. Lett. A
\textbf{54},
225--226
(1975).
\url{https://doi.org/10.1016/0375-9601(75)90174-7}


\bibitem{Gould2004}
C. Gould, C. R\"{u}ster, T. Jungwirth, E. Girgis, G. M. Schott, R. Giraud, K. Brunner, G. Schmidt, and L. W. Molenkamp,
\textit{Tunneling Anisotropic Magnetoresistance: A Spin-Valve-LikeTunnel Magnetoresistance Using a Single Magnetic Layer.}
Phys. Rev. Lett.
\textbf{93},
117203
(2004).
\url{https://doi.org/10.1103/PhysRevLett.93.117203}


\bibitem{Hanneken2015}
C. Hanneken, F. Otte, A. Kubetzka, B. Dup\'{e}, N. Romming, K. von Bergmann, R. Wiesendanger, and S. Heinze,
\textit{Electrical detection of magnetic skyrmions by tunnelling non-collinear magnetoresistance.}
Nature Nanotechnology
\textbf{10},
1039--1042
(2015).
\url{https://doi.org/10.1038/nnano.2015.218}


\bibitem{Kubetzka2017}
A. Kubetzka, C. Hanneken, R. Wiesendanger, and K. von Bergmann,
\textit{Impact of the skyrmion spin texture on magnetoresistance.}
Phys. Rev. B
\textbf{95},
104433
(2017).
\url{https://doi.org/10.1103/PhysRevB.95.104433}


\bibitem{Hayami2021}
S. Hayami and Y. Motome,
\textit{Charge density waves in multiple-Q spin states.}
Phys. Rev. B
\textbf{104},
144404
(2021).
\url{https://doi.org/10.1103/PhysRevB.104.144404}


\bibitem{Wood2023}
G. D. A. Wood, D. D. Khalyavin, D. A. Mayoh, J. Bouaziz, A. E. Hall, S. J. R. Holt, F. Orlandi, P. Manuel, S. Bl\"{u}gel, J. B. Staunton, O. A. Petrenko, M. R. Lees, and G. Balakrishnan,
\textit{Double-Q ground state with topological charge stripes in the centrosymmetric skyrmion candidate GdRu$_{2}$Si$_{2}$.}
Phys. Rev. B
\textbf{107},
L180402
(2023).
\url{https://doi.org/10.1103/PhysRevB.107.L180402}


\bibitem{Sarkar2025}
S. Sarkar, R. Pathak, A. Mukherjee, A. Delin, O. Eriksson, and V. Borisov,
\textit{Magnetic exchange and dipolar interactions in GdRu$_{2}$Si$_{2}$: Three-dimensional magnetism in a layered magnet.}
Phys. Rev. B
\textbf{112},
144414
(2025).
\url{https://doi.org/10.1103/q853-plvr}


\bibitem{Hanaguri2006}
T. Hanaguri,
\textit{Development of high-field STM and its application to the study on magnetically tuned criticality in Sr$_{3}$Ru$_{2}$O$_{7}$.}
J. Phys. Conf. Ser.
\textbf{51},
514
(2006).
\url{https://doi.org/10.1088/1742-6596/51/1/117}


\bibitem{Kohsaka2007}
Y. Kohsaka, C. Taylor, K. Fujita, A. Schmidt, C. Lupien, T. Hanaguri, M. Azuma, M. Takano, H. Eisaki, H. Takagi, S. Uchida, and J. C. Davis,
\textit{An intrinsic bond-centered electronic glass with unidirectional domains in underdoped cuprates.}
Science
\textbf{315}, 
1380--1385
(2007).
\url{https://doi.org/10.1126/science.1138584}


\bibitem{Macdonald2016}
A. J. Macdonald, Y.-S. Tremblay-Johnston, S. Grothe, S. Chi, P. Dosanjh, S. Johnston, and S. A. Burke,
\textit{Dispersing artifacts in FT-STS: a comparison of set point effects across acquisition modes.}
Nanotechnology
\textbf{27},
414004
(2016).
\url{doi:10.1088/0957-4484/27/41/414004}


\bibitem{Thyng2016}
K. M. Thyng, C. A. Greene, R. D. Hetland, H. M. Zimmerle, and S. F. DiMarco,
\textit{True Colors of Oceanography: Guidelines for Effective and Accurate Colormap Selection.}
Oceanography
\textbf{29},
9--13
(2016). 
\url{https://doi.org/10.5670/oceanog.2016.66}


\end{thebibliography}
\end{document}

\title{Supplementary Information for\\`Intra-unit-cell resolved intertwining of multi-\textit{Q} charge and spin textures\\in an itinerant skyrmion magnet'}

% Christopher J. Butler
\author{Christopher J. Butler}
\email{christopher.butler@riken.jp}
\affiliation{RIKEN Center for Emergent Matter Science, Wako, Saitama 351-0198, Japan}

% Katsuki Nihongi
\author{Katsuki Nihongi}
\affiliation{RIKEN Center for Emergent Matter Science, Wako, Saitama 351-0198, Japan}

% Haruto Yoshimochi
\author{Haruto Yoshimochi}
\affiliation{RIKEN Center for Emergent Matter Science, Wako, Saitama 351-0198, Japan}

% Nguyen Duy Khanh
\author{Nguyen Duy Khanh}
\affiliation{Department of Applied Physics and Quantum-Phase Electronics Center (QPEC), University of Tokyo, Tokyo 113-8656, Japan}

% Rina Takagi
\author{\\Rina Takagi}
\affiliation{Institute for Solid State Physics, University of Tokyo, Chiba 277-8581, Japan}

% Tetsuo Hanaguri
\author{Tetsuo Hanaguri}
\email{hanaguri@riken.jp}
\affiliation{RIKEN Center for Emergent Matter Science, Wako, Saitama 351-0198, Japan}

% Shinichiro Seki
\author{Shinichiro Seki}
\affiliation{Department of Applied Physics and Quantum-Phase Electronics Center (QPEC), University of Tokyo, Tokyo 113-8656, Japan}
\affiliation{Research Center for Advanced Science and Technology, University of Tokyo, Tokyo 153-8904, Japan}

\maketitle

\section{Preprocessing of $L(\mathbf{r})$ images for fast Fourier transform}

After cleavage of each GdRu$_{2}$Ge$_{2}$ sample, there is a significant amount of debris on its Ge-terminated surfaces, which cannot be avoided for images larger than a few nm squared. It appears as patches of low intensity in the $50 \times \SI{50}{nm^{2}}$ $L(\mathbf{r})$ maps shown in Fig. 3 of the main article. In fast Fourier transform (FFT) images $\mathcal{F} \left[ L(\mathbf{r}) \right]$, the impact is a significant low-$q$ background that obscures meaningful signals at $\mathbf{Q}_{\mathrm{A}(\mathrm{A}')}$ and $\mathbf{Q}_{\mathrm{B}(\mathrm{B}')}$. To mitigate this, before performing the FFT to each $L(\mathbf{r})$ map we apply a mask created by referring to the simultaneously acquired topography, following the procedure outlined in Fig. S1. Figure S1(a) shows the raw topography. First we subtract an approximately logarithmic contribution to the apparent topographic height originating from piezo creep, and the result of this is shown in Fig. S1(b). We create a preliminary mask image using a chosen height threshold, shown in Fig. S1(c), followed by a Gaussian convolution to slightly expand the zero-valued islands and smooth out their hard edges, as shown in Fig. S1(d). Finally, we use the mask to blend the measured $L(\mathbf{r})$ image with its median value, using $L_{\mathrm{masked}}(\mathbf{r}) = M(\mathbf{r})L(\mathbf{r}) + [1-M(\mathbf{r})]\mathrm{med}[L(\mathbf{r})]$. 

A raw $L(\mathbf{r})$ map, in this case $L(\mathbf{r}, V=\SI{-100}{mV})$ for Phase~III, and its FFT are shown in Figs. S1(e) and S1(f). The image $L_{\mathrm{masked}}(\mathbf{r})$ and its FFT are shown in Figs. S1(g) and S1(h). The Fourier components at $\mathbf{Q}_{\mathrm{A}(\mathrm{A}')}$ and $\mathbf{Q}_{\mathrm{B}(\mathrm{B}')}$ are rendered more visible. The series of $\mathcal{F} \left[ L(\mathbf{r}) \right]$ images displayed in Fig. 3(b) of the main article are preprocessed in this way, using the simultaneously acquired topography image for each respective measurement.

\begin{figure*}
\centering
\includegraphics[scale=1]{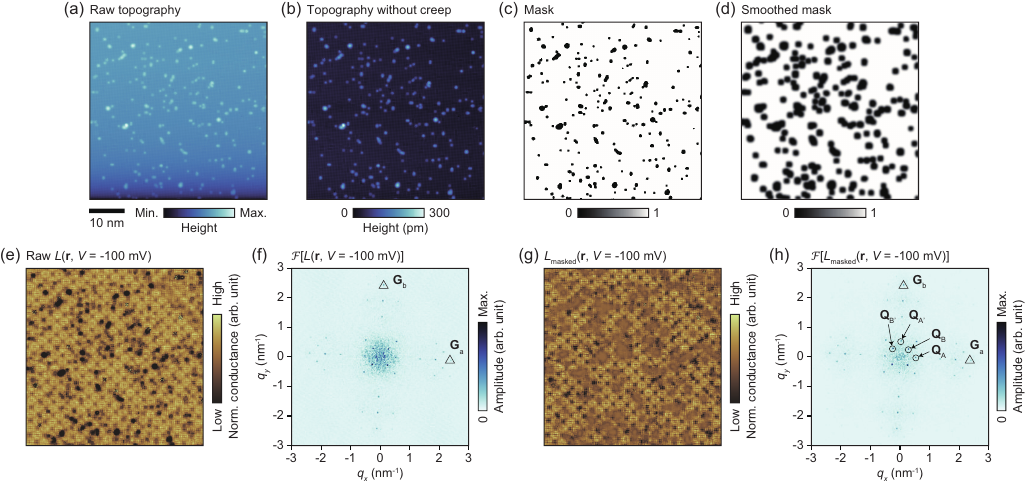}
\caption{\label{fig:1}
\textbf{Preprocessing of $L(\mathbf{r})$ images for FFT.} (a) Raw topography acquired simultaneously with the $L(\mathbf{r}, V=\SI{-100}{mV})$ image for Phase~III and, (b), the topography after subtracting a background that is logarithmic in time, originating from scanner creep. (c) Boolean mask obtained from the topography in (b) using a height threshold. (d) The mask after smoothing with a Gaussian filter and rescaling to the interval $\left[ 0,1 \right]$. (e) Raw $L(\mathbf{r})$ image as presented in Fig. 3(a) of the main article and, (f), its FFT. (g) The masked image and, (h), its FFT, which appears in Fig. 3(b) of the main article.}
\end{figure*}

\section{(In)commensurability of magnetic modulations}

The spatial resolution of scanning tunnelling microscopy (STM) allows an opportunity to investigate single-domain behavior of the charge and spin modulations, and to asses whether they exhibit commensurate lock-in with the crystal lattice, or are instead incommensurate. The latter might generally be expected if the modulation wavevectors are set purely by Fermi surface nesting.

The modulation periodicities are expressed in reciprocal lattice units as $q_{1} = \mathbf{Q}_{\mathrm{A}} / \mathbf{G}_{a}$ and $q_{2} = \mathbf{Q}_{\mathrm{A'}} / \mathbf{G}_{b}$. We estimate their values from the FFT images shown in Fig. 3 of the main article. The wavevectors $\mathbf{Q}_{\mathrm{A}(\mathrm{A'})}$ do not appear in the FFTs for all magnetic phases. Therefore, we use two-dimensional Gaussian fitting to the $\mathbf{G}_{a}\!-\!2\mathbf{Q}_{\mathrm{A}}$ and $\mathbf{G}_{b}\!-\!2\mathbf{Q}_{\mathrm{A'}}$ satellite peaks, which do appear for all magnetic phases, as well as to the Bragg peaks at $\mathbf{G}_{a}$ and $\mathbf{G}_{b}$.

The results, presented in Table~I for each magnetic phase, indicate that the modulations generally deviate slightly from commensurability. The closest commensurate value would be $2/9 = 0.\bar{2}$. These results are consistent with a Fermi surface nesting scenario.

\begin{table}
\caption{Magnetic modulation periodicities estimated from the $\mathcal{F} \left[ L(\mathbf{r}) \right]$ images shown in Fig. 3 of the main article. Error ranges reflect the estimated $\sigma$ values of the relevant Gaussian fitting results when combined using appropriate error propagation.}
\begin{ruledtabular}
\begin{tabular}{ccc}
Magnetic phase & $q_{1}$ (r.l.u.) & $q_{2}$ (r.l.u.) \\
\hline
I   &  0.212 ($\pm0.003$)  &  0.216 ($\pm0.003$) \\
II  &  0.214 ($\pm0.003$)  &  0.217 ($\pm0.003$) \\
III &  0.219 ($\pm0.003$)  &  0.223 ($\pm0.003$) \\
IV  &  0.219 ($\pm0.003$)  &  0.220 ($\pm0.003$) \\
V   &  0.220 ($\pm0.003$)  &  0.220 ($\pm0.003$) \\
\end{tabular}
\end{ruledtabular}
\end{table}

\section{Non-uniform discrete Fourier transform}

\subsection{Motivation}

Here we describe an application of the Fourier transform, other than the FFT, that can be used to neatly describe any observation of a multi-$Q$ periodic scalar field. The basic motivation is that the intra-supercell motif of a multi-$Q$ pattern is determined largely by the relative phases of its Fourier components. For a single- or double-$Q$ scalar field, a change of the phases is equivalent to a change of coordinates and is not interesting. However, as soon as there is a third Fourier component, the new phase degree-of-freedom controls the motif of the pattern. For multi-$Q$ patterns (three components or more), phase degrees-of-freedom can control the pattern's symmetry including its chirality and chiral handedness, for example, and it can be important to precisely measure multiple phases for a periodic pattern appearing in an STM or other experiment. Even for a simpler double-$Q$ pattern such as that of a commensurate charge density wave (CDW), the inclusion of the Bragg peaks alongside the CDW peaks renders the system `multi-$Q$' in the sense that the phase degrees-of-freedom can describe intra-supercell details such as the atomic-scale registry between the CDW and the lattice.

\begin{figure*}
\centering
\includegraphics[scale=1]{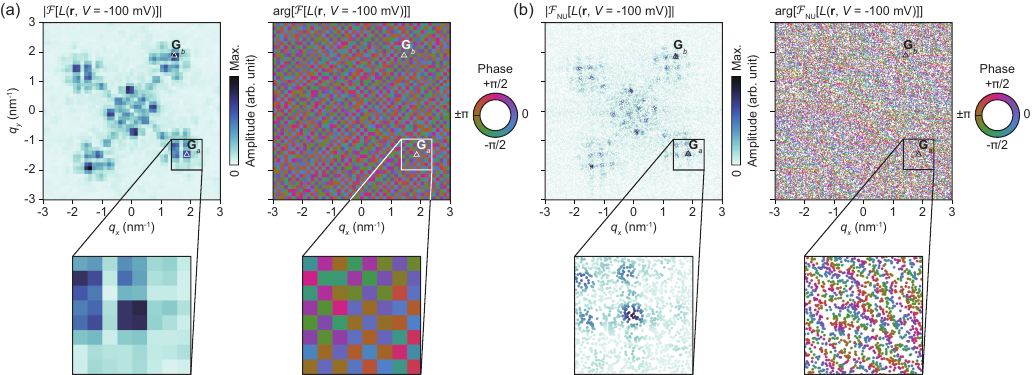}
\caption{\label{fig:2}
\textbf{Demonstration of phase measurement at arbitrary $\mathbf{Q}$ using the NUDFT.} (a) Amplitude and phase images obtained using the usual FFT applied to a representative $L(\mathbf{r})$ image [for Phase~II, from Fig. 4(a) of the main article]. Zooming in around $\mathbf{G}_{a}$ shows that a precise location and amplitude for this peak (or any other) can be estimated with a suitable fitting procedure, but the same procedure will not yield a useful value from the phase image. While the FFT here has very low $q$-space resolution due to the small $r$-space field-of-view, this low resolution is not the real problem. Increasing it does nothing to obviate the basic restrictions associated with the FFT. (b) Amplitude and phase of Fourier coefficients $c_{m}$ obtained at $5 \times 10^{4}$ points randomly sampled throughout the same $q$-space field-of-view as for (a) using the NUDFT.}
\end{figure*}

Measuring the phase of more than two modulations cannot generally be done using the FFT because it restricts the sampling of $q$-space to a regular grid, and Fourier components of real patterns generally do not fall on this grid, so are shared over multiple pixels.
%The FFT gives complex-valued output and therefore a complete description of the input image (though the phase information is almost always discarded before visualization). 
While it is straightforward to use two-dimensional peak fitting to estimate a more precise $q$-space location and amplitude for a peak in the FFT, it is much less straightforward to retrieve phase information from across multiple pixels in a similar way. (It is still possible, but requires a procedure more complicated that the one described below.) This problem is clearly seen in Fig. S2(a), where we show the FFT-derived amplitude and phase images of a representative $L(\mathbf{r})$ map. Use of the Lawler-Fujita algorithm \cite{Hytch1998,Lawler2010} can reduce the width for \textit{at most two} FFT peaks down to a single pixel each, thereby giving access to useful phase information for those two. However, if there are more than two important peaks this is not sufficient.

The desired multi-$Q$ phase information can be obtained using a non-uniform discrete Fourier transform (NUDFT), in this case of Type-I, or `NUDFT-I' (going from a uniform $r$-space sampling to a non-uniform $q$-space sampling). This is demonstrated in Fig. S2(b) where, alongside the usual FFT result $\mathcal{F}[L(\mathbf{r})]$, we show the corresponding NUDFT result $\mathcal{F}_{\mathrm{NU}}[L(\mathbf{r})]$ obtained at $5 \times 10^{4}$ points randomly sampled throughout the same $q$-space field-of-view. With this we are free to sample $q$-space precisely where it is required, and obtain unambiguous phase information.

To express the above idea in a more concrete form, we specify the following: For some quantity $X(\mathbf{r})$, a uniform grid sampling in $r$-space yields $N$ measured values $X_{n}$ at measurement locations $\mathbf{r}_{n}$. This is the output from our microscope, for example. Given this, we can obtain a single complex Fourier coefficient for each of a collection of $M$ arbitrary $q$-space locations $\mathbf{Q}_{m}$:
$$
c_{m} = \frac{1}{N} \sum_{n}^{N} X_{n} e^{-2 \pi i \mathbf{Q}_{m} \cdot \mathbf{r}_{n}}.
$$
Obtaining coefficients $c_{m}$ for uniformly spaced points in $q$-space would yield the same result as the FFT (but less quickly). Instead we target a non-uniformly distributed set of points $\mathbf{Q}_{m}$, which we refer to collectively as $\mathcal{Q}$. The resulting collection of Fourier coefficients is our selective Fourier spectrum for an image, $\mathcal{F}_{\mathcal{Q}}[X(\mathbf{r})]$.
%(not to be confused with the usual Fourier transform $\mathcal{F}[X(\mathbf{r})]$).

\subsection{Target Q vectors for NUDFT}

In the present case, the set of vectors in $q$-space that we expect to be physically meaningful include, at minimum, the magnetic modulation vectors 
$$
\{ \mathbf{Q}_{\mathrm{A}}, \mathbf{Q}_{\mathrm{B}}, \mathbf{Q}_{\mathrm{A}'}, \mathbf{Q}_{\mathrm{B}'} \}.
$$
Given some generic magnetic modulation with wavevector $ \mathbf{Q}_{\mathrm{mag.}}$, if there is any spin-charge coupling the charge distribution may acquire a modulation at $2 \mathbf{Q}_{\mathrm{mag.}}$. This is because the spin-charge coupling should be independent of the local moments' directions but might depend on their magnitude which, if it is modulated at all, is modulated at twice the wavevector $ \mathbf{Q}_{\mathrm{mag.}}$. Therefore, in the charge sector we may expect to see the $2^{\mathrm{nd}}$ harmonics
$$
\{ 2 \mathbf{Q}_{\mathrm{A}}, 2 \mathbf{Q}_{\mathrm{B}}, 2 \mathbf{Q}_{\mathrm{A}'}, 2 \mathbf{Q}_{\mathrm{B}'} \}.
$$

\begin{figure}
\centering
\includegraphics[scale=1]{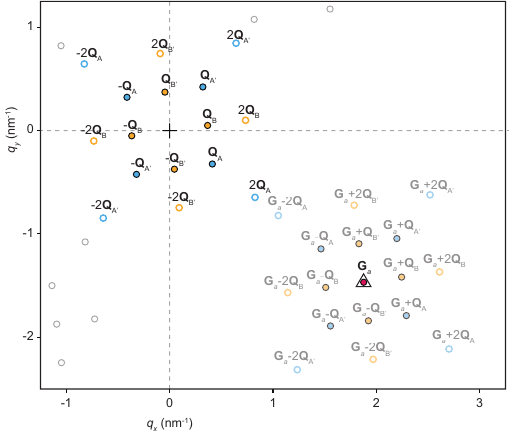}
\caption{\label{fig:3}
\textbf{Target $\mathbf{Q}$ vectors for NUDFT.} A subset of the collection of wavevectors $\mathcal{Q}$, including the primary magnetic wavevectors, their $2^{\mathrm{nd}}$ harmonics, the Bragg vector $\mathbf{G}_{a}$, and the satellites of $\mathbf{G}_{a}$.
}
\end{figure}

Because our STM observations resolve the atomic lattice, we can also address the Bragg vectors
$$
\{ \mathbf{G}_{a}, \mathbf{G}_{b} \}.
$$
Finally, we also see satellite peaks around the Bragg peaks [see Fig. 3(b) of the main article]. These are linear combinations of vectors from the sets listed above. For example, the set of satellites around $\mathbf{G}_{a}$ is
$$
\{ \mathbf{G}_{a} \pm \mathbf{Q}_{\mathrm{A}}, \mathbf{G}_{a} \pm \mathbf{Q}_{\mathrm{B}}, \ldots \mathbf{G}_{a} \pm 2 \mathbf{Q}_{\mathrm{B}'} \}.
$$
A physical interpretation of these satellites is that, together with each Bragg peak, they form beat patterns that envelope a feature that would otherwise be distributed with atomic periodicity, such as particular orbitals arrayed on the lattice. Therefore, they describe phenomena such as modulated bond/orbital order. We will return to this when discussing the `basketweave' pattern in Sec.~V below. Henceforth, $\mathcal{Q}$ will denote the complete set of interesting $q$-space vectors, the union of the sets listed above, i.e.:
$$
\mathcal{Q} = \{ \mathbf{G}_{a}, \mathbf{G}_{b}, \mathbf{Q}_{\mathrm{A}}, \mathbf{Q}_{\mathrm{B}}, \ldots \mathbf{G}_{b} \pm 2 \mathbf{Q}_{\mathrm{B}'} \}.
$$
This has 84 members in total, but only 42 need to be addressed because the Fourier coefficient measured at any wavevector $\mathbf{Q}_{m}$ is the complex conjugate of its counterpart at $-\mathbf{Q}_{m}$. A part of their arrangement is depicted in Fig. S3.

%We target only the $q$ points on the positive-$q_{x}$ side of $q$ space, which encompasses the unique information. (Data at, for example, $-\mathbf{Q}_{\mathrm{A}}$ on the negative-$q_{x}$ side is the complex conjugate of data at $\mathbf{Q}_{\mathrm{A}}$ on the positive side.)

One way of using the NUDFT to acquire phases at the wavevectors in $\mathcal{Q}$ would be to first acquire precise $q$-space locations using a two-dimensional Gaussian fit to each peak in the FFT amplitude image, and then perform the NUDFT at those precise wavevectors. For a one-off treatment of a single image this would be fine, but in the present case we use a different procedure. We need to process a series of up to 70 $L(\mathbf{r})$ maps in which particular Fourier components appear and disappear with changes in magnetic field. We require a procedure that i) enables a high degree of automation, and ii) treats images acquired in different magnetic phases on an equal footing by virtue of being robust against the vanishing of amplitude for any subset of $\mathcal{Q}$. The work-flow is as follows:
\begin{enumerate}

\item Perform a two-dimensional Gaussian fit to the Bragg peaks in the simultaneously acquired topography, obtaining provisional coordinates for $\mathbf{G}_{a}$ and $\mathbf{G}_{b}$.

\item Make an initial guess for the complete set of $q$-space vectors $\mathcal{Q}$ using the estimates of $q_{1}$ and $q_{2}$ listed in Table~I.

\item Improve upon this initial guess using a Markov chain Monte Carlo-inspired gradient-ascent algorithm. This works as follows:

\begin{enumerate}
\item Take a small Gaussian random step in the six-dimensional space formed by $G_{a,x}$, $G_{a,y}$, $G_{b,x}$, $G_{b,y}$, $q_{1}$ and $q_{2}$,

\item Generate a new set of wavevectors for this iteration, $\mathcal{Q}_{i}^{\mathrm{iter.}}$, and use the NUDFT on the given $L(\mathbf{r})$ map to measure the sum of the Fourier amplitudes for $\mathcal{Q}_{i}^{\mathrm{iter.}}$. If this total amplitude increases, keep the new $\mathcal{Q}_{i}^{\mathrm{iter.}}$. Otherwise reject the step and revert to the previous configuration $\mathcal{Q}_{i-1}^{\mathrm{iter.}}$.

\item Repeat.

\end{enumerate}

This algorithm runs for a fixed length of 2000 steps.

\item Having arrived close to the configuration of $\mathcal{Q}$ that maximizes the total Fourier amplitude, return the final Fourier spectrum $\mathcal{F}_{\mathcal{Q}}[L(\mathbf{r})]$.

\item Move on to the next $L(\mathbf{r})$ image in the series and repeat.

\end{enumerate}
Results of this procedure applied to a single measurement, i.e. for both the simultaneously acquired topography, $\mathcal{F}_{\mathcal{Q}}[Z(\mathbf{r})]$ and for the conductance, $\mathcal{F}_{\mathcal{Q}}[L(\mathbf{r})]$, are shown in Figs. S4(a) and S4(b).

\subsection{Phase-referencing of Fourier spectra $\mathcal{F}_{\mathcal{Q}}[L(\mathbf{r})]$}

Figures S4(a) and S4(b) show NUDFT results for simultaneously acquired topographic and $L(\mathbf{r})$ maps, i.e. $\mathcal{F}_{\mathcal{Q}}[Z(\mathbf{r})]$ and $\mathcal{F}_{\mathcal{Q}}[L(\mathbf{r})]$, respectively. In Fig. S4(b), there is no immediately obvious structure among the various Fourier components' phases. This is because all the phases are referenced against the origin in $r$-space, which is at the lower-left corner of the measurement's field-of-view and is therefore arbitrary. Moreover, when we change the magnetic field by a small amount and repeat the measurement in the same magnetic phase and field-of-view (precise to $\sim \SI{1}{\AA}$), a different configuration of phases appears. This is because of a coherent sliding of the spin structure, and therefore its LDOS imprint, which occurs when the field is changed.

\begin{figure*}
\centering
\includegraphics[scale=1]{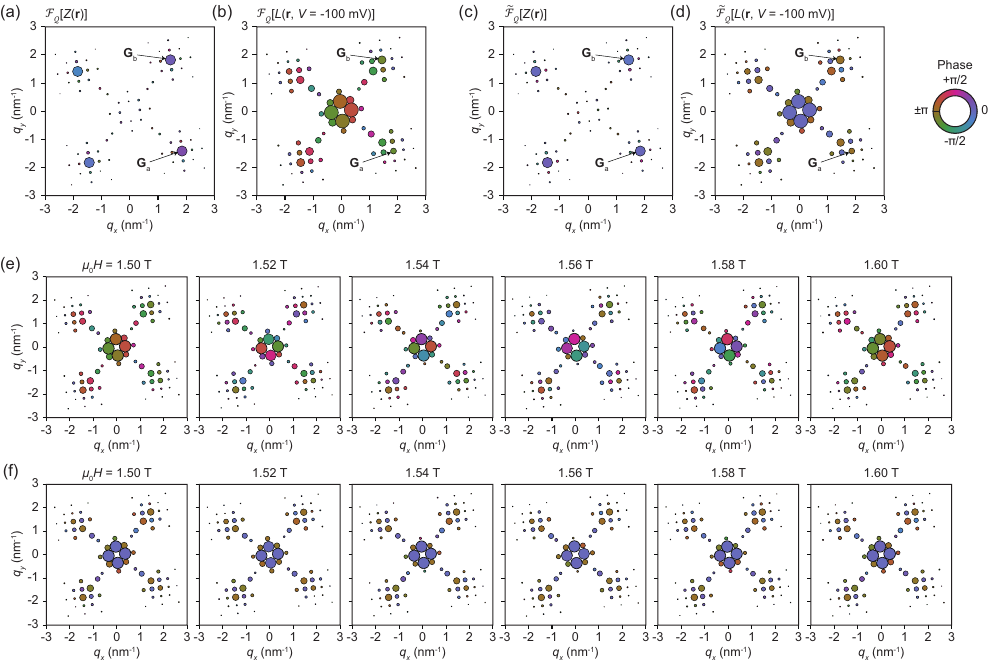}
\caption{\label{fig:4}
\textbf{Phase-referencing of Fourier spectra $\mathcal{F}_{\mathcal{Q}}[L(\mathbf{r})]$.} (a) and (b) Examples of raw NUDFT results for simultaneously measured topography and $L(\mathbf{r}, V=\SI{-100}{mV})$ images. (c) and (d) The same data after the phase-referencing procedure. Before the procedure, it can be seen that the Bragg components in $Z(\mathbf{r})$ and $L(\mathbf{r})$ are anti-phase with each other, and this is clearer after phase-referencing.
%The phases measured at the Bragg vectors for the $L$ image are rotated by $\pi$ with respect to those for the topography image. This is because the intensity in $L$ images is, on average, hollow-centred with respect to the Ge and Gd lattices [see Fig. 5(b) of the main article for an example].
(e) Raw $\mathcal{F}_{\mathcal{Q}}[L(\mathbf{r})]$ results for images selected from the measurement series underpinning the analysis described in Fig. 4 of the main article. The results displayed here are all acquired within Phase~IV, and therefore describe similar LDOS modulations. (f) The same series of NUDFT results after applying the phase-referencing procedure. The data in (e) and (f) encode exactly the same series of $r$-space patterns (albeit with descriptions stemming from different $r$-space origins), but those in (f) are more easily interpretable and comparable with each other.
}
\end{figure*}

If the origin happened to coincide with the corner of the magnetic unit-cell, the internal structure of the set of phases would become clear. We now describe a \textit{phase-referencing} procedure to realize this situation, finding and then applying $r$-space translations such that the uninteresting phase offsets are removed. We do this on a measurement-by-measurement basis. For any given measurement the location $\mathbf{r}$ of a data point can be decomposed into $\mathbf{r}_{0}$, a translation that takes us from the origin point of the given measurement to a corner of the magnetic unit-cell, and $\mathbf{r}_{\mathrm{intra.}}$, which is then the location of the data point within the magnetic unit-cell. We would like to have an intra-unit-cell version of the general expression for a Fourier coefficient $c_{m}$ given above, i.e.:
$$
c_{\mathrm{intra.},m} = \frac{1}{N} \sum_{n}^{N} X_{n} e^{-2 \pi i \mathbf{Q}_{m} \cdot \mathbf{r}_{\mathrm{intra.},n} }.
$$
To get there we can substitute $\mathbf{r}\!=\!\mathbf{r}_{0}\!+\!\mathbf{r}_{\mathrm{intra.}}$ and pull out a $q$-dependent phase factor that encodes the effect of the translation by vector $\mathbf{r}_{0}$:
\begin{align*}
c_{m} &= \frac{1}{N} \sum_{n}^{N} X_{n} e^{-2 \pi i \mathbf{Q}_{m} \cdot ( \mathbf{r}_{0} + \mathbf{r}_{\mathrm{intra.}, n} ) } \\
c_{m} &= e^{-2 \pi i \mathbf{Q}_{m} \cdot \mathbf{r}_{0} } \frac{1}{N} \sum_{n}^{N} X_{n} e^{-2 \pi i \mathbf{Q}_{m} \cdot \mathbf{r}_{\mathrm{intra.},n} }.
\end{align*}
We can multiply both sides by $e^{2 \pi i \mathbf{Q}_{m} \cdot \mathbf{r}_{0}}$ and then define 
$$
c_{\mathrm{intra.},m} = e^{2 \pi i \mathbf{Q}_{m} \cdot \mathbf{r}_{0}} c_{m},$$
which is our desired phase-referenced Fourier coefficient.

How do we find a translation vector $\mathbf{r}_{0}$ that sets the reference point to a corner of the magnetic unit-cell? Since its size is set by $\mathbf{Q}_{\mathrm{B}}$ and $\mathbf{Q}_{\mathrm{B'}}$, there is only one location per unit-cell where the phases of these modulations are both zero, i.e. $\mathrm{arg}[c_{\mathrm{intra.},\mathbf{Q}_{\mathrm{B}}}] = \mathrm{arg}[c_{\mathrm{intra.},\mathbf{Q}_{\mathrm{B'}}}] = 0$, and so this is a sensible choice for the corner location. Practically, to realize these conditions we perform a brute-force grid search of an auxiliary space of possible translations, and select the one that causes $c_{\mathrm{intra.},m}$ to fulfill the above criteria. (For the images acquired in Phase~II, $\mathbf{Q}_{\mathrm{B}(\mathrm{B}')}$ are effectively absent, so instead we use $\mathbf{G}_{a}\!+\!\mathbf{Q}_{\mathrm{B'}}$ and $\mathbf{G}_{a}\!-\!\mathbf{Q}_{\mathrm{B}}$, which in concert with the following criteria leads to an equivalent result -- also see Sec.~V on the Fourier components of the basketweave pattern.)

We also impose that the $r$-space translation sets the phases for both Bragg wavevectors to $\pi$, i.e. $\mathrm{arg}[c_{\mathrm{intra.},\mathbf{G}_{a}}] = \mathrm{arg}[c_{\mathrm{intra.},\mathbf{G}_{b}}] = \pi$. By itself this corresponds to positioning the reference point on a particular Ge ion somewhere, or equivalently, a Gd ion. All these criteria can be fulfilled simultaneously, given a sufficiently wide-ranging search of the space of translations. This is because no two of the wavevectors $\mathbf{G}_{a(b)}$ and $\mathbf{Q}_{\mathrm{B}(\mathrm{B}')}$ have projections on each other that are commensurate with either one. Simply, somewhere there is a Gd ion located exactly on the corner of the magnetic unit-cell.

\begin{figure*}
\centering
\includegraphics[scale=1]{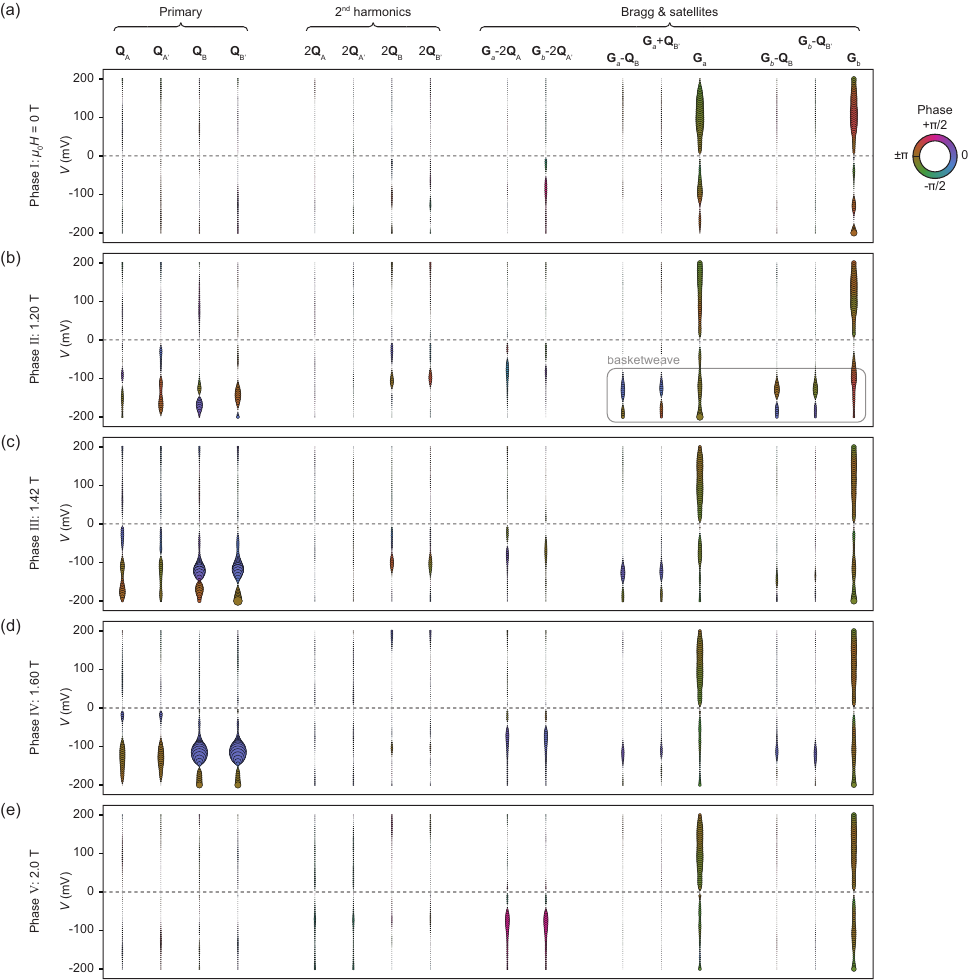}
\caption{\label{fig:5}
\textbf{Energy-dependence of Fourier coefficients.} (a)--(e), Amplitudes and phases for a selected subset of $\tilde{\mathcal{F}}_{\mathcal{Q}}[L(\mathbf{r}, V)]$ results in the bias interval $V = 200$ to $\SI{-200}{mV}$, for Phases~I--V, respectively. The $L(\mathbf{r}, V)$ datasets underpinning these results were acquired in the same field-of-view as for those shown in Fig. 4 of the main article, in a separate magnetic field sweep ($V_{\mathrm{set}} = \SI{200}{mV}$, $I_{\mathrm{set}} = \SI{200}{pA}$, $V_{\mathrm{mod}} = \SI{5}{mV}$). Here we present the amplitudes with a different visual scaling as compared to other plots: The diameter of each marker, rather than the area, corresponds to the amplitude. In panel (b), the components that encode the basketweave pattern are highlighted.
}
\end{figure*}

Requiring that the phases of the Bragg modulations of $L(\mathbf{r})$ should be set to $\pi$ is equivalent to requiring that the phases of the Bragg modulations of the topography should be 0. This is because the intensity in the $L(\mathbf{r})$ images is, on average, hollow-centred (i.e. in anti-phase) with respect to the Ge lattice seen in $Z(\mathbf{r})$. This relationship is clearly visible in $r$-space in Fig. 5(b) of the main article, and can also be seen in $q$-space as shown in Fig. S4. Here, it is already clear for the raw NUDFT result shown for each type of image in Figs. S4(a) and S4(b), but is clearer for the phase-referenced NUDFT result in Figs. S4(c) and S4(d).

The overall result of the above procedure is that the phases in our NUDFT results are referenced against the corner of the magnetic unit cell and simultaneously against the location of a single Gd ion. Its effectiveness can be seen by comparing the raw and phase-referenced NUDFT results for $L(\mathbf{r})$ maps acquired in a short magnetic field interval. Raw $\mathcal{F}_{\mathcal{Q}}[L(\mathbf{r})]$ results for such a series are displayed in Fig. S4(e), and the phase-referenced results, now denoted as $\tilde{\mathcal{F}}_{\mathcal{Q}}[L(\mathbf{r})]$, are displayed in Fig. S4(f).

%As an aside, if an exhaustive search of `translation space' does not yield a translation that causes the phase relations in two data sets to match, this indicates that the two measured patterns are meaningfully different, even if that difference is not apparent to the eye.

\section{Energy-dependence of Fourier coefficients}

Throughout the main article we show images $L(\mathbf{r}, V=\SI{-100}{mV})$, close to the bias voltage where the LDOS modulations in each magnetic phase are seen to be most prominent. Here, in Fig. S5, we show the energy-dependence of selected Fourier coefficients in the interval $V = 200$ to $\SI{-200}{mV}$, measured in the same field-of-view as for the data shown in Fig. 4 of the main article. The Fourier coefficients are phase-referenced following the same procedure described in Sec.~III,C above, using the $L(\mathbf{r}, V=\SI{-100}{mV})$ image extracted from each measurement. From Fig. S5, we see that the common feature for all energies in all magnetic phases is that the phase of the Bragg wavevectors is consistently close to $\pi$, albeit with small fluctuations that might be related to the details of the apex orbital of the STM tip. This indicates that the LDOS is consistently hollow-centred with respect to the Gd square-net over the whole energy range investigated here.

Aside from the Bragg wavevectors, all Fourier components for all magnetic phases have small intensities for $V > \SI{0}{mV}$. The Fourier components that encode the characteristic patterns in each magnetic phase have significant intensity in the range $V = -100$~to~$\SI{-150}{mV}$, and the $\tilde{\mathcal{F}}_{\mathcal{Q}}[L(\mathbf{r}, V=\SI{-100}{mV})]$ plots shown in Fig. 4(b) of the main article are representative of these most intense patterns. In conclusion, although the Gd spin-spin dot product model presented in the main article does not take energy into account, in the energy range where the LDOS modulations are found to be most prominent, the patterns' qualitative features are successfully captured by the model.

\section{Fourier components of the basketweave pattern}

\begin{figure}
\centering
\includegraphics[scale=1]{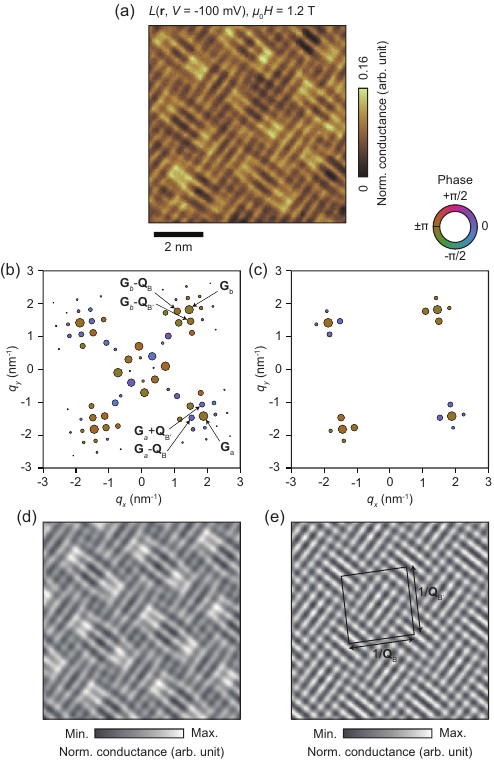}
\caption{\label{fig:6}
\textbf{Fourier components of the basketweave pattern.} (a) An $L(\mathbf{r})$ map acquired in Phase~II, reproduced from Fig. 4 of the main article. (b) The corresponding $\tilde{\mathcal{F}}_{\mathcal{Q}}[L(\mathbf{r})]$ plot and, (c), a plot of the subset of Fourier components responsible for generating the basketweave motif. (d) A plot of the plane waves of $\tilde{\mathcal{F}}_{\mathcal{Q}}[L(\mathbf{r})]$ back into $r$-space. This is analogous to an inverse Fourier transform back to the source image. The apparent sliding of the pattern is due to the phase-referencing procedure described above. (e) The corresponding plot taking only the Bragg and satellite components shown in (c). Note that it is the two-fold symmetry of the phase configuration of the satellites in (c), and not any difference in the components' amplitudes, that generates the two-fold symmetry of the basketweave motif.
}
\end{figure}

The $L(\mathbf{r})$ images acquired in Phase~II show a striking basketweave pattern with several interesting features. First, it is composed of two interwoven bond-order-like stripe components, each of which is spatially modulated in anti-phase with the other. Second, it is clear to the eye that the pattern has a supercell size of $1 / \mathbf{Q}_{\mathrm{B}}\!\times\!1 /\mathbf{Q}_{\mathrm{B}'}$, but the $\mathcal{F}[L(\mathbf{r})]$ and $\tilde{\mathcal{F}}_{\mathcal{Q}}[L(\mathbf{r})]$ data show negligible Fourier components at $\mathbf{Q}_{\mathrm{B}}$ and $\mathbf{Q}_{\mathrm{B}'}$. The supercell size is instead set by a combination of other Fourier components, namely the ones describing the apparent modulated bond order, the Bragg modulations their nearby satellites.
%As mentioned above, the Bragg peaks and satellites can together form a beat pattern that envelopes a feature that would otherwise exhibit atomic periodicity. Therefore, they describe phenomena such as modulated bond/orbital order.

\begin{figure*}
\centering
\includegraphics[scale=1]{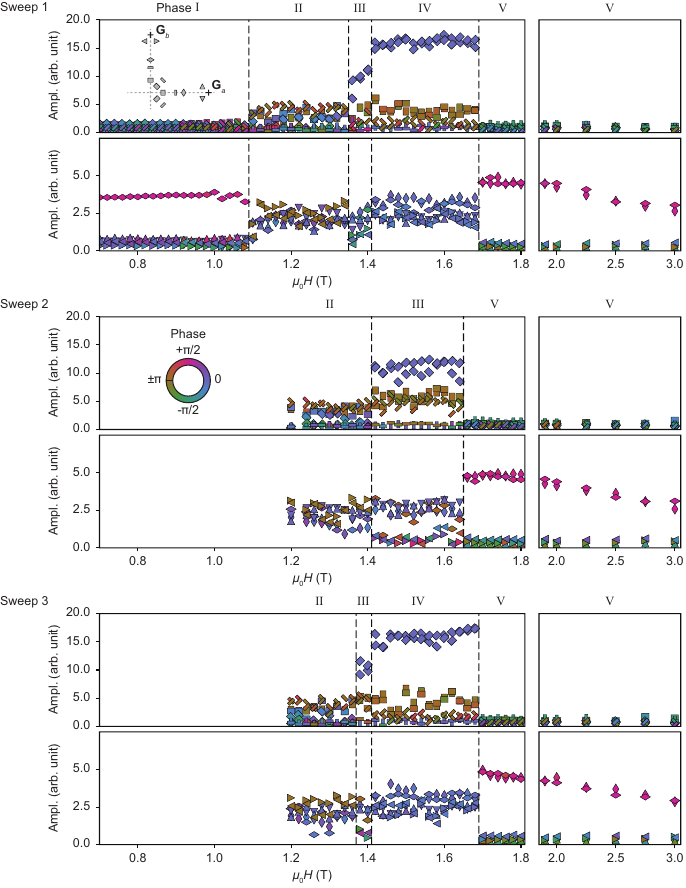}
\caption{\label{fig:7}
\textbf{Field-dependence of Fourier coefficients upon repetition of magnetic field sweep.} Here we track a selected subset of $\tilde{\mathcal{F}}_{\mathcal{Q}}[L(\mathbf{r}, V=\SI{-100}{mV})]$ results for images acquired in three magnetic field sweeps. Between each sweep, the field was ramped to $\mu_{0}H = \SI{12}{T}$ to fully polarize the sample. Magnetic phase transitions are indicated by vertical dashed lines. In each sweep the phase transitions, at least in this microscopic field-of-view, are seen to occur at different field strengths, and in the second sweep there is a transition directly from Phase~III to V, skipping Phase~IV.}
\end{figure*}

In Fig. S6 we show how the basketweave pattern is generated by the interference of the lattice corrugations described by the Bragg wavevectors, $\mathbf{G}_{a(b)}$, and modulations described by their satellites, $\mathbf{G}_{a}\!-\!\mathbf{Q}_{\mathrm{B}}$, etc. Figures S6(a) and S6(b) show the observed pattern and its Fourier spectrum, as reproduced from Fig. 4 of the main article. Figure S6(c) shows the subset of Fourier components which are found to encode the key properties of the basketweave pattern when plotted as plane waves back in $r$-space. Note that it is the specific phase relations among these Fourier components, and not their amplitudes, that encode the peculiar two-fold rotationally symmetric motif. A consideration of Figs. S6(c) and S6(e) suggests the following: If the supercell is chosen as shown in Fig. S6(e), and the reference point for the phases is the corner of the supercell, then the reference point is at a location where the bond-order-like arrangement is most intense (either along one bond axis or the other). Here, the Bragg and satellite modulations oriented along one bond axis each have a phase of $\sim \pi$, and interfere constructively. Along the perpendicular axis, however, the Bragg modulation has phase $\pi$ but the satellites have phase $0$, so they interfere destructively. Shifting the supercell by $(\pm 1 / 2\mathbf{Q}_{\mathrm{B}'}, \pm 1 / 2\mathbf{Q}_{\mathrm{B}})$ would invert these relationships: The phase structure in Fig. S6(c) would appear to rotate by $90^{\circ}$.

An interesting phenomenon associated with the basketweave pattern can be seen in the energy-dependence results shown in Fig. S5. At an energy of approximately $\SI{-150}{meV}$, the phases of the satellites $\mathbf{G}_{a}\!-\!\mathbf{Q}_{\mathrm{B}}$, etc., all invert, but the phases of the Bragg peaks do not. This means that the bond-order-like pattern changes its orientation by $90^{\circ}$ as the energy changes, in a way reminiscent of some electronic nematic systems \cite{Butler2022}.

\section{Repetition of magnetic field sweep}

Figure S7 shows subsets of the $\tilde{\mathcal{F}}_{\mathcal{Q}}[L(\mathbf{r})]$ results acquired throughout the interval $ \mu_{0} H = 0.7$ to $\SI{3.0}{T}$. The measurements were performed in three sweeps, and between each sweep the field was increased to $\SI{12}{T}$ so that the sample was fully polarized and any magnetic `memory' was erased. In each sweep, the magnetic phase transitions are observed to occur at different field strengths. Moreover, in Sweep~2, the magnetic state changes directly from Phase~III to V, skipping Phase~IV. These variations are a consequence of the $1^{\mathrm{st}}$ order nature of the phase transitions.

\begin{table*}
\caption{Magnetic structure model parameters reproduced from \cite{Yoshimochi2024}, but here including the revised ground state}
\begin{ruledtabular}
\begin{tabular}{cccccc}
Magnetic phase  &  $m_{c}^{0}$  & $\nu$ & $\mathbf{Q}_{\nu}$                                                    &  $\tilde{m}(\mathbf{Q}_{\nu})$ & $\theta^{\mathbf{Q}_{\nu}}$ \\
\hline
I               &       0       &   1   &  ($q_{\mathrm{I},1}$, 0, 0)                                           &  (1, 0, 1)                     & (0, 0, $\frac{\pi}{2}$) \\
\hline
I (rev.)        &       0       &   1   &  ($q_{\mathrm{I},1}$, 0, 0)                                           &  (1, 0, 1)                     & (0, 0, $\frac{\pi}{2}$) \\
                &               &   2   &  (0, $q_{\mathrm{I},2}$, 0)                                           &  (0.02, 0.02, 0.5)             & ($\pi$, $\frac{\pi}{2}$, $\frac{3\pi}{2}$) \\
\hline
II              &     0.23      &   1   &  ($q_{\mathrm{II},1}$, 0, 0)                                          &  (0.89, 0, 1)                  & ($\frac{3\pi}{2}$, 0, $\pi$) \\
                &               &   2   &  (0, $q_{\mathrm{II},2}$, 0)                                          &  (0.24, 0, 0.59)               & ($\frac{3\pi}{2}$, 0, $\pi$) \\
                &               &   3   &  ( $\frac{q_{\mathrm{II},1}}{2}$,  $\frac{q_{\mathrm{II},2}}{2}$, 0)  &  (0.59, 0, 0.39)               & (0, 0, $\frac{3\pi}{2}$)     \\
                &               &   4   &  ($-\frac{q_{\mathrm{II},1}}{2}$, $\frac{q_{\mathrm{II},2}}{2}$, 0)   &  (0.59, 0, 0.39)               & (0, 0, $\frac{3\pi}{2}$)      \\
                &               &   5   &  ( $q_{\mathrm{II},1}$, $q_{\mathrm{II},2}$, 0)                       &  (0.15, 0, 0.08)               & ($\frac{3\pi}{2}$, 0, $\pi$) \\
                &               &   6   &  ($-q_{\mathrm{II},1}$, $q_{\mathrm{II},2}$, 0)                       &  (0.15, 0, 0.08)               & ($\frac{3\pi}{2}$, 0, $\pi$) \\
\hline
III             &     0.41      &   1   &  ($q_{\mathrm{III},1}$, 0, 0)                                         &  (0.35, 0.26, 0.58)            & ($\frac{3\pi}{2}$, $\frac{3\pi}{2}$, $\pi$) \\
                &               &   2   &  (0, $q_{\mathrm{III},2}$, 0)                                         &  (0.12, 0, 0.38)               & ($\frac{3\pi}{2}$, 0, $\frac{3\pi}{2}$) \\
                &               &   3   &  ( $\frac{q_{\mathrm{III},1}}{2}$, $\frac{q_{\mathrm{III},2}}{2}$, 0) &  (0.46, 0.08, 0.11)            & ($\frac{\pi}{2}$, 0, 0) \\
                &               &   4   &  ($-\frac{q_{\mathrm{III},1}}{2}$, $\frac{q_{\mathrm{III},2}}{2}$, 0) &  (0.46, 0.08, 0.11)            & ($\frac{\pi}{2}$, $\pi$, 0) \\
\hline
IV              &     0.57      &   1   &  ($q_{\mathrm{IV},1}$, 0, 0)                                          &  (0.18, 0, 0.43)               & ($\frac{3\pi}{2}$, 0, $\pi$) \\
                &               &   2   &  (0, $q_{\mathrm{IV},2}$, 0)                                          &  (0.18, 0, 0.43)               & ($\frac{3\pi}{2}$, 0, $\pi$) \\
                &               &   3   &  ($ \frac{q_{\mathrm{IV},1}}{2}$, $\frac{q_{\mathrm{IV},2}}{2}$, 0)   &  (0.30, 0, 0.15)               & ($\frac{\pi}{2}$, 0, 0) \\
                &               &   4   &  ($-\frac{q_{\mathrm{IV},1}}{2}$, $\frac{q_{\mathrm{IV},2}}{2}$, 0)   &  (0.30, 0, 0.15)               & ($\frac{\pi}{2}$, 0, 0) \\
                &               &   5   &  ($q_{\mathrm{IV},1}$, $q_{\mathrm{IV},2}$, 0)                        &  (0.07, 0, 0.03)               & ($\frac{3\pi}{2}$, 0, $\pi$) \\
                &               &   6   &  ($-q_{\mathrm{IV},1}$, $q_{\mathrm{IV},2}$, 0)                       &  (0.07, 0, 0.03)               & ($\frac{3\pi}{2}$, 0, $\pi$) \\
\hline
V               &     0.72      &   1   &  ($q_{\mathrm{V},1}$, 0, 0)                                           &  (0, 0, 0.48)                  & (0, 0, $\frac{\pi}{2}$) \\
                &               &   2   &  (0, $q_{\mathrm{V},2}$, 0)                                           &  (0, 0, 0.48)                  & (0, 0, $\frac{\pi}{2}$) \\
\end{tabular}
\end{ruledtabular}
\end{table*}

These measurements were acquired with an increasing magnetic field, and it is reasonable to ask whether hysteresis can be observed for the downward sweep. However, in the present case, a straightforward observation of magnetic hysteresis should not be expected. Hysteresis is an aggregate behavior of an ensemble of many microscopic subdivisions of a sample, only one of which can be monitored using the present measurement scheme.
%The reader is invited to consider whether hysteresis can be meaningfully observed in an Ising or Potts model simulation, if the observation is restricted to a single pixel or small cluster of pixels.

\section{Spin structure models,\\ incl. for revised ground state}

\subsection{Recipe for representing multi-\textit{Q} spin structures}

To obtain the $W(\mathbf{r})$ images presented in the main article, we first need to model the spin structure for each magnetic phase. For this we follow the recipe described by Yoshimochi \textit{et al.} \cite{Yoshimochi2024} which we briefly review here. Each magnetic structure is represented as the interference of periodic modulations of the local magnetic moment. Each modulation is represented as
\begin{equation*}
\mathbf{m}(\mathbf{r}) = \sum_{ \alpha } \mathbf{e}_{\alpha} \mathrm{sin} (\mathbf{Q} \cdot \mathbf{r} + \theta_{\alpha})
\end{equation*}
where $\mathbf{Q}$ is the propagation wavevector. The index $\alpha$ runs over axes of a reference frame oriented around the propagation direction: The axis parallel to the crystal lattice vector $\mathbf{c}$, the one parallel to $\mathbf{Q}$, and the one parallel to $\mathbf{c}\!\times\!\mathbf{Q}$, respectively. The components of $\mathbf{m}(\mathbf{r})$ in this frame are each modulated sinusoidally with respective phase offsets given by the phase vector $\theta_{\alpha}$. For example, a spin screw is described with $\theta_{\alpha} = (\theta_{\mathbf{c}}, \theta_{\mathbf{Q}}, \theta_{\mathbf{c}\!\times\!\mathbf{Q}}) =(0, 0, \pi / 2)$, setting the two components that are perpendicular to the propagation vector a quarter of a wavelength out of phase with each other.

The superposition of multiple modulations, with an additional term for a uniform $c$-axis magnetization under external field, $m^{0}_{c}$, is written as
\begin{equation*}
\mathbf{m}(\mathbf{r}) = \mathbf{e}_{c} m^{0}_{c} + \sum_{ \alpha, \nu} \mathbf{e}_{\alpha} \tilde{m}_{\alpha}(\mathbf{Q}_{\nu}) \mathrm{sin} (\mathbf{Q}_{\nu} \cdot \mathbf{r} + \theta^{\mathbf{Q}_{\nu}}_{\alpha}).
\end{equation*}
In this multi-$Q$ case the components of the various phase vectors $\theta_{\alpha}^{\mathbf{Q}_{\nu}}$ have a second important role: Controlling the detailed motif within each magnetic unit-cell, which can in turn control properties such as the magnetic unit-cell's topological index \cite{Hayami2021b} (see Sec.~VIII).

The parameters used to represent all magnetic structures considered in this work are displayed in Table~II.

\begin{figure*}
\centering
\includegraphics[scale=1]{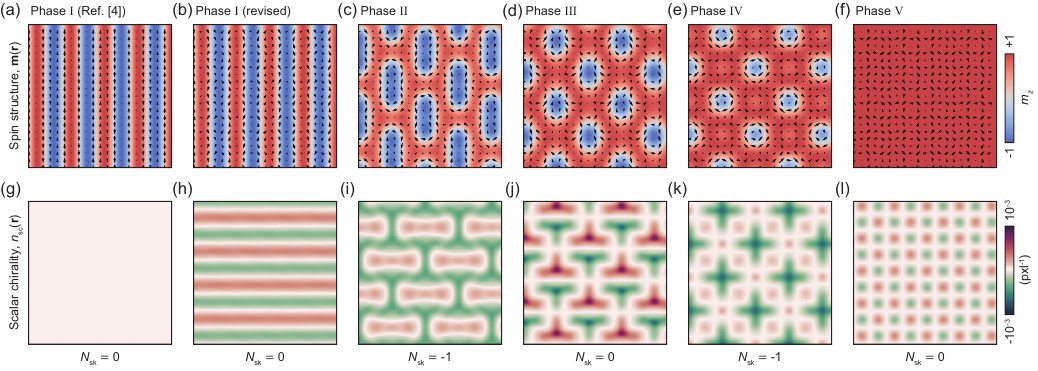}
\caption{\label{fig:8}
\textbf{Scalar chirality distributions for various spin textures.} (a) Depiction of the single-$Q$ screw spin structure proposed for the ground state by Yoshimochi \textit{et al.} \cite{Yoshimochi2024}. (b) The revised, double-$q$ ground state spin structure. (c)--(f), Spin structures for Phases~II--V. (g)--(l), The corresponding scalar chirality images. The value of $N_{\mathrm{sk}}$ computed for each structure is given below each panel. The scalar image for the revised ground state, (h), exhibits stripes similar to those recently proposed for the ground state of GdRu$_{2}$Si$_{2}$ \cite{Khanh2022,Wood2023,Spethmann2024}, though the integrated scalar chirality remains zero.
}
\end{figure*}

\subsection{Multi-\textit{Q} ground state}

In Fig. 3(b) of the main article, we find that there are Fourier components in the $L(\mathbf{r})$ map which involve both $\mathbf{Q}_{\mathrm{A}}$ and $\mathbf{Q}_{\mathrm{B}}$, which signifies straight away that it is not a single-$Q$ structure as initially thought \cite{Yoshimochi2024}. In fact, the $W(\mathbf{r})$ image calculated for the single-$Q$ screw structure does not exhibit a Fourier component even at $\mathbf{Q}_{\mathrm{A}}$. This is because dot products between nearest-neighbor spins are spatially uniform in a screw structure, and the outcome in $W(\mathbf{r})$ is a bond-order-like pattern with wavevector $\mathbf{G}_{b}$, and no low-$Q$ modulations. (This is concretely demonstrated in Figs. S8 and S9, below.)

Through a process of trial-and-error, we find that it is possible to model the `anomalous' $L(\mathbf{r})$ modulations highlighted in Fig. 3(b) by adding only one extra magnetic modulation to our representation of the ground state, and then calculating its $W(\mathbf{r})$ image. This additional modulation has wavevector $\mathbf{Q}_{\mathrm{A'}}$ and is described in full detail in Table II. Surprisingly, Fourier components of $W(\mathbf{r})$ that involve $\mathbf{Q}_{\mathrm{B}}$ can be generated even without adding any $\mathbf{Q}_{\mathrm{B}}$ modulations to our representation of the magnetic structure. The manifestation of these extra modulations in $W(\mathbf{r})$, and a correct prediction for the overall symmetry of the Fourier pattern, are achieved by adjusting the differences between the components of $\theta_{\alpha}^{\nu=2}$. Furthermore, a good prediction of the phases of the measured Fourier coefficients $\tilde{c}_{m}$ [Fig. 6(c) of the main article] is achieved by adjusting the difference between $\theta_{\alpha}^{\nu=2}$ and $\theta_{\alpha}^{\nu=1}$, while preserving the internal configuration of each.

Our revised model for the ground state is the outcome of an \textit{ad hoc} trial-and-error approach, but if the approach can be systematized it might be useful in future investigations. Any method of predicting an $L(\mathbf{r})$ image from some spin structure $\mathbf{m}(\mathbf{r})$ can in principle be used to `learn' a real spin structure through an iterative process of modifying $\mathbf{m}(\mathbf{r})$ until a good prediction is achieved.

\section{Scalar chirality maps,\\ incl. for revised ground state}

For a vector field such as that embodied by local magnetic moments $\mathbf{m}(\mathbf{r})$, a scalar chirality can be calculated as
\begin{equation*}
n_{\mathrm{sc}}(\mathbf{r}) = \mathbf{n}(\mathbf{r}) \cdot \left[ \frac{\partial \mathbf{n}(\mathbf{r})}{\partial x} \times \frac{\partial \mathbf{n}(\mathbf{r})}{\partial y} \right] ,
\end{equation*}
where $\mathbf{n}(\mathbf{r}) = \mathbf{m}(\mathbf{r})/ \left| \mathbf{m}(\mathbf{r}) \right| $. For a periodic configuration, this scalar chirality can be integrated over the unit-cell to yield a topological index which takes on integer values, the skyrmion number 
\begin{equation*}
N_{\mathrm{sk}} = \frac{1}{4 \pi} \int n_{\mathrm{sc}}(\mathbf{r}) \; d^{2} \mathbf{r}.
\end{equation*}
For a skyrmion phase, $N_{\mathrm{sk}}~=~-1$, and for trivial phases such as screw, helical, meron-antimeron lattice, and vortex lattice phases, $N_{\mathrm{sk}}~=~0$. It is directly associated with magneto-transport properties measurable in macroscopic samples, such as the topological Hall effect \cite{Taguchi2001a,Taguchi2001b,Tokura2020}.

Figure S8 displays the spin structures $\mathbf{m}(\mathbf{r})$ considered in this work and generated as described in Sec.~VII above, along with their respective scalar chirality maps, $n_{\mathrm{sc}}(\mathbf{r})$. In Fig. S8(b) we include the revised model of the ground state spin structure and in Fig. S8(h) we include its scalar chirality map. Here we see scalar chirality stripes, or `topological charge stripes' very similar to those in the recently proposed ground state spin structure of GdRu$_{2}$Si$_{2}$ \cite{Khanh2022,Wood2023,Spethmann2024}. However, the global topological index for this structure, as for the single-$Q$ screw phase, is zero.

\section{Shortcomings of $W(\mathbf{r})$ maps as predictions of observed $L(\mathbf{r})$}

\begin{figure*}
\centering
\includegraphics[scale=1]{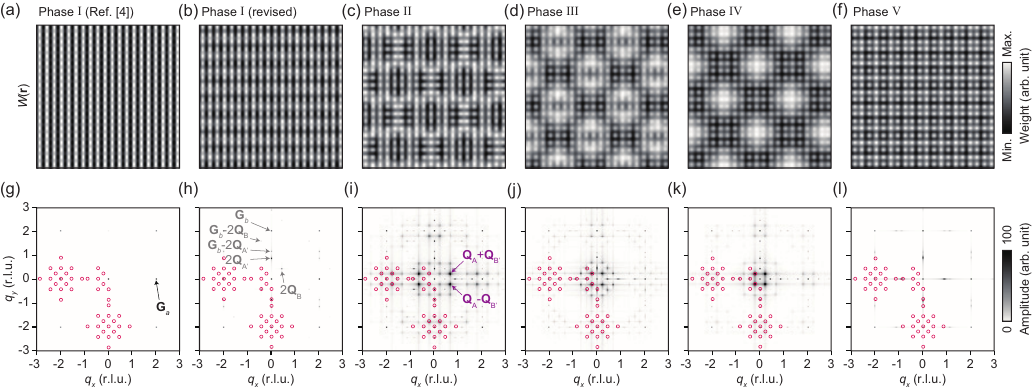}
\caption{\label{fig:9}
\textbf{Calculated $W(\mathbf{r})$ and $\mathcal{F}[W(\mathbf{r})]$ images for various spin textures.} (a)--(f), $W(\mathbf{r})$ images for the same spin structures plotted in Fig. S8 above. The field-of-view for these images is $20\!\times\!20$ lattice units. (g)--(l), FFT images for each respective pattern for a field-of-view of $48\!\times\!48$ lattice units. The set of wavevectors $\mathcal{Q}$ that we address in the rest of this work are marked with red circles. In (h), the Fourier components that emerge upon revising the ground state as described above, and which are seen in measurements [Fig. 3(b) of the main text] are marked with gray labels. In (i), prominent Fourier components of the $W(\mathbf{r})$ image that do not appear in the measurement, of the type $ \mathbf{Q} = \pm n\mathbf{Q}_{\mathrm{A}(\mathrm{A'})}\!\pm\!m\mathbf{Q}_{\mathrm{B}(\mathrm{B'})} $, are marked with purple labels.
}
\end{figure*}

In the main article we describe a scheme for predicting the observed $L(\mathbf{r})$ maps. We construct $W(\mathbf{r})$ maps based on the calculation of nearest-neighbor spin-spin dot products for the spin structures described above, followed by convolution of these bond-centred dot product results with kernel functions that represent the contribution of each given bond to the LDOS. This scheme successfully captures several qualitative features of the observations. For example, it correctly predicts the tendency of the modulations at $\mathbf{Q}_{\mathrm{B}(\mathrm{B'})}$ to increases in intensity upon transition from Phase~II to III, and again from III to IV, before drastically diminishing in Phase~V. Notably, it also successfully predicts the configuration of phases for those Fourier components that encode the basketweave pattern, described above. 

However, it also has clear shortcomings. Some of these can be seen in the FFT images, $\mathcal{F}[W(\mathbf{r})]$, shown in Fig. S9. The most notable mismatch with observation is the appearance of Fourier components at wavevectors that are linear combinations of $\mathbf{Q}_{\mathrm{A}(\mathrm{A'})}$ and $\mathbf{Q}_{\mathrm{B}(\mathrm{B'})}$, which are not included in the set of target wavevectors described in Sec.~III,B. These do not appear in observations at $V = \SI{-100}{mV}$ (see Fig. 3 of the main article) and only a small subset of them appear, only very faintly, at other energies. As indicated in Fig. S9(i), these Fourier components are most prominent in the $W(\mathbf{r})$ map for Phase~II. From a comparison of Figs. 4(b) and 6(c) of the main article, we also see that the prediction of the relative amplitudes and phases of the low-$q$ modulations that are common to both observed $L(\mathbf{r})$ and predicted $W(\mathbf{r})$, (e.g. $\mathbf{Q}_{\mathrm{A}}, \mathbf{Q}_{\mathrm{B}}, 2\mathbf{Q}_{\mathrm{A}}, 2\mathbf{Q}_{\mathrm{B}}$,...) are the worst for Phase~II. It is reasonable to infer that these two facts are somehow related.

Possible explanations for these discrepancies include:
\begin{enumerate}
\item The spin structures underpinning the construction of $W(\mathbf{r})$ maps are inaccurate,
\item The dot product scheme does not fully capture the physics controlling the LDOS weight at each bond (certainly true),
\item The kernel functions that represent the contribution to the LDOS at each bond are poorly designed.
\end{enumerate}
Of these possibilities, the third can be easily dismissed: As long as the kernel functions remain realistically simple, they cannot generate new Fourier components in the convolution result, or extinguish existing ones. (Any alterations to the kernel functions only have the effect of rotating the phases of the Fourier components that would already have existed if the kernels had been delta functions, as per the convolution theorem.) Therefore the source of the additional components must be somewhere other than in the form of the kernels. Most likely, the spin-spin dot product approach does not capture the relevant physics determining the LDOS, which is not surprising given its extreme simplicity.